\documentclass[a4paper,fleqn]{cas-dc}

\usepackage[authoryear]{natbib}

\usepackage{amsmath, amsfonts, amsthm, bbm, mathtools,bm,float,graphicx,placeins,caption} 

\def\tsc#1{\csdef{#1}{\textsc{\lowercase{#1}}\xspace}}
\tsc{WGM}
\tsc{QE}
\tsc{EP}
\tsc{PMS}
\tsc{BEC}
\tsc{DE}
\begin{document}
\let\WriteBookmarks\relax
\def\floatpagepagefraction{1}
\def\textpagefraction{.001}

% Short title
\shorttitle{Energy Dissipation in Synchronous Binary Asteroids}

% Short author
\shortauthors{A. J. Meyer et al.}

% Main title of the paper
\title [mode = title]{Energy Dissipation in Synchronous Binary Asteroids}                       
% Title footnote mark
% eg: \tnotemark[1]

% Title footnote 1.
% eg: \tnotetext[1]{Title footnote text}
% \tnotetext[<tnote number>]{<tnote text>} 

\author[1]{Alex J. Meyer}[orcid=0000-0001-8437-1076]
\cormark[1]
\address[1]{Smead Department of Aerospace Engineering, University of Colorado, Boulder, CO 80303, USA}

\author[1]{Daniel J. Scheeres}[]

\author[2]{Harrison F Agrusa}[orcid=0000-0002-3544-298X]
\address[2]{Department of Astronomy, University of Maryland, College Park, MD 20742, USA}

\author[3]{Guillaume Noiset}[orcid=0000-0002-1649-7176]
\address[3]{Royal Observatory of Belgium, 3 Avenue Circulaire, 1180 Brussels, Belgium}

\author[1]{Jay McMahon}[]

\author[3]{{\"O}zg{\"u}r Karatekin}[]

\author[4,5]{Masatoshi Hirabayashi}[orcid=0000-0002-1821-5689]
\address[4]{Department of Aerospace Engineering, Auburn University, Auburn, AL 36849, USA}
\address[5]{Department of Geosciences, Auburn University, Auburn, AL 36849, USA}

\author[4]{Ryota Nakano}[orcid=0000-0002-9840-2416]

\cortext[1]{Corresponding author at  3775 Discovery Dr, Boulder, CO 80303, USA\\ E-mail address: alex.meyer@colorado.edu}

% Here goes the abstract
\begin{abstract}
\noindent{Synchronous binary asteroids can experience libration about their tidally-locked equilibrium, which will result in energy dissipation. This is an important topic to the Asteroid Impact and Deflection Assessment, where excitation caused by the DART kinetic impact in the Didymos binary asteroid system may be reduced through dissipation before Hera arrives to survey the effects of the impact. We develop a numeric model for energy dissipation in binary asteroids to explore how different system configurations affect the rate of energy dissipation. We find tumbling within the synchronous state eliminates a systematic trend in libration damping on short timescales (several years), but not over long times (hundreds of years) depending on the material conditions. Furthermore, damping of libration, eccentricity, and fluctuations in the semimajor axis are primarily dependent on the stiffness of the secondary, whereas the semimajor axis secular expansion rate is dictated by the stiffness of the primary, as expected. Systems experiencing stable planar libration in the secondary can see a noticeable reduction in libration amplitude after only a few years depending on the stiffness of the secondary, and thus dissipation should be considered during Hera's survey of Didymos. For a very dissipative secondary undergoing stable libration, Hera may be able to calculate the rate of libration damping in Dimorphos and therefore constrain its tidal parameters.} 
\end{abstract}

% Use if graphical abstract is present
% \begin{graphicalabstract}
% \includegraphics{figs/grabs.pdf}
% \end{graphicalabstract}

% Research highlights
% \begin{highlights}
% \item We develop a numeric model to simulate energy dissipation in a librating synchronous binary asteroid.
% \item Eccentricity damps much quicker in binary asteroids than predicted by analytic tidal theory.
% \item The dissipation rate of libration amplitude, eccentricity, and semimajor axis fluctuations are coupled and driven by the tidal parameters of the secondary.
% \item Tumbling in the synchronous state damps at roughly the same rate as stable libration.
% \item The libration amplitude of Dimorphos after the DART impact may noticeably decrease before the arrival of Hera.
% \end{highlights}

% Keywords
% Each keyword is seperated by \sep

\maketitle

\section{Introduction} \label{sec:intro} %###################################################
%###################################################%###################################################
The Asteroid Impact and Deflection Assessment (AIDA) is a collaboration supported by NASA and ESA to test the feasibility of a kinetic impactor to deflect a small asteroid for the purpose of planetary defense \citep{cheng2018aida}. Two missions will combine results to produce the most accurate knowledge possible on the first kinetic impact of an asteroid: NASA's DART (Double Asteroid Redirection Test), which will perform the actual kinetic impact \citep{rivkin2021double}, and ESA's Hera, which will assess the effectiveness of the impact several years later \citep{michel2022esa}. The target of the impact is Dimorphos, the secondary in the Didymos binary asteroid system. By impacting Dimorphos, DART will change the mutual orbit period around Didymos, and ground-based measurements of the orbit period change will reveal how much momentum was transferred to Dimorphos. Approximately 4 years after the DART impact, Hera is scheduled to rendezvous with the Didymos system to perform a detailed analysis of the post-impact system, making several key measurements. 

The degree to which the system's dynamics will evolve through energy dissipation between the DART impact and Hera's arrival remains an open question for AIDA. While this window is only around 4 years, a rubble-pile structure -- like Didymos is hypothesized to be based on earlier dynamics and geological studies \citep{agrusa2022dynamical,walsh2018rubble,walsh2008rotational,jacobson2011dynamics} -- may be very efficient at dissipating energy, and thus this problem warrants attention. The question of energy dissipation after the DART impact and prior to Hera's rendezvous with Didymos is necessary in order to maximize the scientific and practical return of the AIDA collaboration. As Hera characterizes the spin state of Dimorphos, it is important to understand how the current spin state has changed since the impact in order to fully comprehend the effects of the DART impact. Ignoring dissipation in the system may lead to an incorrect estimation of the efficacy of a kinetic impactor during Hera's survey. While the scientific implications of this work extend beyond AIDA and Didymos to binary asteroid dynamics in general, we focus our analysis to this application given its current relevance and the wealth of analysis on Didymos and the DART impact in the literature.

While binary asteroids provide an ideal test site for planetary defense missions given their short mutual orbit periods \citep{cheng2018aida}, they also offer a chance to study the unique dynamics of the full 2-body problem (F2BP). Given the asteroids' close proximity and generally asymmetric shapes \citep{pravec2016binary}, their orbital motion is strongly coupled with their attitude, leading to complex dynamics \citep{maciejewski1995reduction, scheeres2006relative, scheeres2009stability}. Through this strong coupling, the bodies' spins and mutual orbit will evolve concurrently while energy dissipation occurs. Additionally, spin-orbit coupling can lead to attitude instabilities as a result of orbit perturbations such as the DART impact \citep{agrusa2021excited}.

There are two main mechanisms of energy dissipation we will consider in this work, both stemming from the deformation of the bodies. The first is tidal torque, in which the tidal forces of both bodies act to move the system into a synchronous equilibrium \citep{murray1999solar,goldreich2009tidal, taylor2010tidal}. The second is non-principal axis (NPA) rotation, in which rotation about any axis other than the major principal axis will dissipate energy until the major principal axis is aligned with the angular momentum \citep{burns1973asteroid,breiter2012stress, molina2003energy, ershkov2019dynamics, pravec2005tumbling}. Both these mechanisms will drive the system toward a configuration in which the two asteroids are mutually tidally locked, with their spin angular momentum vectors aligned with their major principal axes and the orbit angular momentum vector \citep{taylor2011binary}. We call this state the doubly-synchronous equilibrium. 

While many studies focus on energy dissipation in the two-body problem, they generally ignore the specific dynamical regime that Didymos will inhabit after the DART impact: a system which is generally synchronous but with nonzero libration of the secondary \citep{taylor2010tidal, goldreich2009tidal}. Here we define libration as any angular deviation of the secondary's long axis away from the tidally locked configuration, but smaller than $90^\circ$ so the secondary remains on-average synchronous. Generally there are two modes of libration: free and forced \citep{murray1999solar}. While forced libration is driven by eccentricity, free libration is governed by the average libration over an orbit period and is thus eccentricity-agnostic \citep{tiscareno2009rotation}. Given the strongly coupled nature of binary asteroids, we make no distinction between these two modes and simply adopt the physical libration angle. This study will focus exclusively on this dynamic state and so also carries scientific merit beyond the specific application of the AIDA collaboration. More recently, \cite{efroimsky2018dissipation} analyzed energy dissipation in a tidally perturbed librating body. This is the same regime we are interested in here, but we attempt to relax the small-libration assumption from that work and extend results to binary asteroids, which orbit much closer than planet-moon systems. Another noteworthy study is that of \cite{jacobson2011dynamics}, who apply a tidal torque model to binary asteroids. However, this analysis is limited to 2 dimensions, whereas we are interested in the full 3 dimensional dynamics. \cite{quillen2020excitation, quillen2022non} study tidal dissipation in coupled systems with some attention spent on the libration state, and our work falls in a similar vein but we focus on how different shapes and stiffness of the secondary affect the dissipation process.

Since the main motivation of this study is the AIDA collaboration, we first provide background on Didymos, the DART impact, and previous analyses on the post-impact dynamics in Section \ref{sec:background}. We then derive our dynamical model, including dissipation mechanisms, in Section \ref{sec:dynamics}. Results on energy dissipation are presented in Section \ref{sec:Dissipation}, and we validate these results by comparing with a higher-fidelity numeric model in Section \ref{sec:gubas}. The implications for Hera over the short-term are investigated in Section \ref{sec:Short Term}. In Section \ref{sec:byorp} we discuss the possible implications of the BYORP effect, and in Section \ref{sec:QK} we investigate how the dissipation behavior depends on the material parameters. Finally, we present a discussion and our conclusions in Section \ref{sec:discussion}.

\section{Background} \label{sec:background} %###################################################
%###################################################%###################################################
We will apply our dissipation model to the Didymos system, which we nominally assume is in a singly-synchronous equilibrium prior to any perturbation, with the secondary's rotation period equal to the orbit period. The rationale for this assumption is outlined in \cite{richardson2022predictions}. To calculate this equilibrium we adopt the method described in \cite{agrusa2021excited} and iterate the system bulk density until the stroboscopic orbit period matches the observed value. We define the stroboscopic orbit period as the time required for the secondary to traverse $360^\circ$ relative to an inertial observer, akin to a lightcurve observation. This approach means we have developed our own independent estimate of the system density rather than using values derived from observations, although our density lies within the error bars of the observed value \citep{naidu2020radar,scheirich2022preimpact}. We calculate the stroboscopic orbit period using the method outlined in \cite{meyer2021libration}. The resulting equilibrium system has the parameters outlined in Table \ref{didymos_parameters}. We will assume a triaxial shape for the secondary, but note changing the axis ratios of Dimorphos does not appreciably affect the equilibrium parameters of the system. While keeping the mean radius of Dimorphos constant, we will vary its axis ratios ($a/b$ and $b/c$, with $a>b>c$) to investigate how the shape of Dimorphos affects the energy dissipation rate.

In this work we will focus on two shapes of the secondary, one with $a/b=1.2$, $b/c=1.1$, and the other with $a/b=1.4$, $b/c=1.3$. In conjunction with the mean radius, we can solve for the semiaxes that define the ellipsoid, as well as the dimensionless shape parameter $S$ defined as
\begin{equation}
    S=\frac{B-A}{C}
\end{equation}
where $A$, $B$, and $C$ are the three principal moments of inertia of the ellipsoid corresponding to the axes $a$, $b$, and $c$, respectively. Table \ref{dimorphos_parameters} gives the dimensions of the two ellipsoids we will primarily use as Dimorphos, as well as their shape parameter $S$.

\begin{table*}[ht!]
\caption{Summary of the equilibrium Didymos system prior to the DART impact, from \cite{pravec2022photometric}, \cite{scheirich2022preimpact}, \cite{naidu2020radar}, and \cite{scheirich2009modeling}. Our density estimate differs from that reported in \cite{scheirich2022preimpact} as we calculate it using a dynamical approach, but our solution falls within the $1\sigma$ derived error bars.}
\begin{center}
\begin{tabular}{| c || c | c | c |}
\hline
 Parameter & Symbol & Value & Notes\\ 
 \hline
 Orbit Period & $P_{orb}$ & 11.92 hr & Measured \citep{pravec2022photometric,scheirich2022preimpact}\\
 Didymos Rotation Period & $P_{A}$ & 2.26 hr & Measured \citep{pravec2022photometric}\\
 Didymos Mean Radius & $R_{A}$ & 390 m & Derived \citep{naidu2020radar}\\
 Dimorphos Mean Radius & $R_{B}$ & 82 m & Derived \citep{naidu2020radar,scheirich2009modeling}\\
 System Bulk Density & $\rho$ & 2.2 g/cm$^3$ & Derived here, similar to \citep{scheirich2022preimpact}\\
 Semimajor Axis & $a$ & 1200 m & Measured \citep{naidu2020radar}\\
 Eccentricity & $e$ & 0 & Assumed \citep{scheirich2022preimpact,richardson2022predictions}\\
 Inclination & $i$ & 0$^\circ$ & Assumed \citep{scheirich2022preimpact,richardson2022predictions}\\
  \hline
  \end{tabular}
  \label{didymos_parameters}
\end{center}
\end{table*}

\begin{table}[ht!]
\caption{Summary of the two triaxial ellipsoids used throughout this work as the secondary shape.}
\begin{center}
\begin{tabular}{| c || c | c | c | c |}
\hline
 Shape & $a$ [m] & $b$ [m] & $c$ [m] & $S$ \\ 
 \hline
 $a/b=1.2$, $b/c=1.1$ & 95.6 & 79.7 & 72.4 & 0.18 \\
 $a/b=1.4$, $b/c=1.3$ & 112 & 80.0 & 61.5 & 0.32 \\
  \hline
  \end{tabular}
  \label{dimorphos_parameters}
\end{center}
\end{table}

The DART impact will push Dimorphos out of the equilibrium state \citep{meyer2021libration,agrusa2021excited}. The impact can be quantified by the momentum enhancement factor known as $\beta$, which is defined as the ratio of the true system momentum change to the momentum carried by the impactor. Mathematically, this is described as
\begin{equation}
    \beta = \frac{p_{true}}{p_{impactor}}.
\end{equation}
$\beta$ can be converted into a change in velocity using the relationship 
\begin{equation}
    \Delta\vec{v} = \frac{M_{impactor}}{M_B}\left(\vec{u}+(\beta-1)\left(\hat{u}\cdot\vec{u}\right)\hat{n}\right)
\end{equation}
where $M_{impactor}$ is the impactor mass, $M_B$ is the mass of Dimorphos, $\vec{u}$ is the impactor velocity, and $\hat{n}$ is the outward surface normal at the impact site \citep{rivkin2021double, feldhacker2017shape}. For this analysis we will assume $\hat{n}$ is parallel to the velocity of Dimorphos, and that $\vec{u}$ is misaligned with the velocity vector by 10 degrees out of the orbit plane and 10 degrees in the radial direction, consistent with \cite{richardson2022predictions}. Note the impact is retrograde, decreasing the mutual orbit period while increasing its eccentricity. From \cite{richardson2022predictions}, we use an impactor mass of 536 kg and  velocity of 6.143 km/s relative to the secondary.

\begin{figure*}[ht!]
   \centering
   \includegraphics[width = \textwidth]{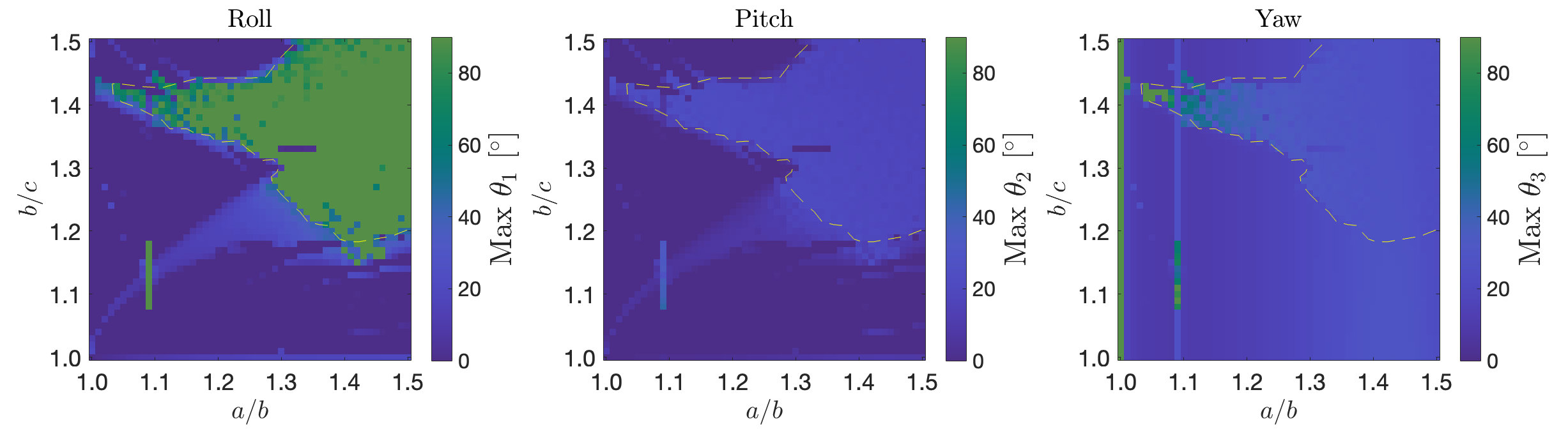} 
   \caption{The maximum amplitude of the 1-2-3 Euler angles for an impact corresponding to $\beta=3$ ($e\approx0.02$), from the simulation set from \cite{agrusa2021excited}. The unstable regions, indicated by nonzero amplitudes in the roll and pitch angles, are governed by intersections of various resonances among fundamental frequencies of the system. The unstable region is outlined by the yellow dashed line; this is not a formal boundary and only serves to aid interpretation.}
   \label{beta3_impact}
\end{figure*}

For this analysis, we will assume a perturbation equivalent to a $\beta$ value of 3, as this is large enough to excite unstable motion in some shapes of Dimorphos, but small enough to allow stable motion in other shapes, and also lies in the expected range of $\beta$: $1< \beta \lesssim6$ \citep{raducan2022global,stickle2022effects}. A perturbation of $\beta=3$ is roughly equivalent to increasing the eccentricity to around 0.02, depending on the secondary shape and mass. This allows us to study both stable and unstable dynamical regimes without having to test multiple perturbation magnitudes. Note this selection is not grounded in the actual DART impact, as we are interested in how different secondary shapes affect dissipation rates within a system rather than making any quantitative predictions, as an accurate prediction is impossible without knowledge of the system's interior structure. Following the DART impact and Hera survey, this analysis can be revisited with better constraints on the shape and mass of Dimorphos. We reproduce the results of \cite{agrusa2021excited} for $\beta=3$ in Fig. \ref{beta3_impact} to show the unstable region of motion. While the size of Dimorphos is fixed by the bulk diameter, the shape of the triaxial ellipsoid is defined by the axis ratios $a/b$ and $b/c$, where $a$, $b$, and $c$ are the longest, intermediate, and shortest semiaxes of the ellipsoid, respectively. Fig. \ref{beta3_impact} shows the amplitude of the 1-2-3 Euler angles, corresponding to roll, pitch, and yaw, for each secondary shape. If a system is in a true equilibrium, these angles would remain zero. An unstable region is apparent in Fig. \ref{beta3_impact} where some secondary shapes result in tumbling. The unstable region in which the secondary begins to tumble is outlined by a yellow dashed line. This is not a formal boundary for this region and is only intended to aid interpretation. The unstable region is dependent on the system's eccentricity, so this boundary can not be applied outside of our impact scenario.

Due to the spin-orbit coupling in binary asteroids, these systems are non-Keplerian, and thus osculating Keplerian elements can be somewhat misleading. In an equilibrium configuration, the secondary may appear to be in a circular orbit to an external observer, but the Keplerian orbit is elliptical. In this configuration, the secondary is trapped at periapsis while the orbit itself precesses. Thus, there is a non-zero eccentricity at equilibrium and the semimajor axis is not the same as the separation distance \citep{scheeres2009stability}. However, these elements are still useful as they can give us an idea of the system's secular evolution over time, and we use the Keplerian osculating elements throughout this work.

\section{Dynamical Model} \label{sec:dynamics} %###################################################
\begin{figure}[ht!]
   \centering
   \includegraphics[width = 3in]{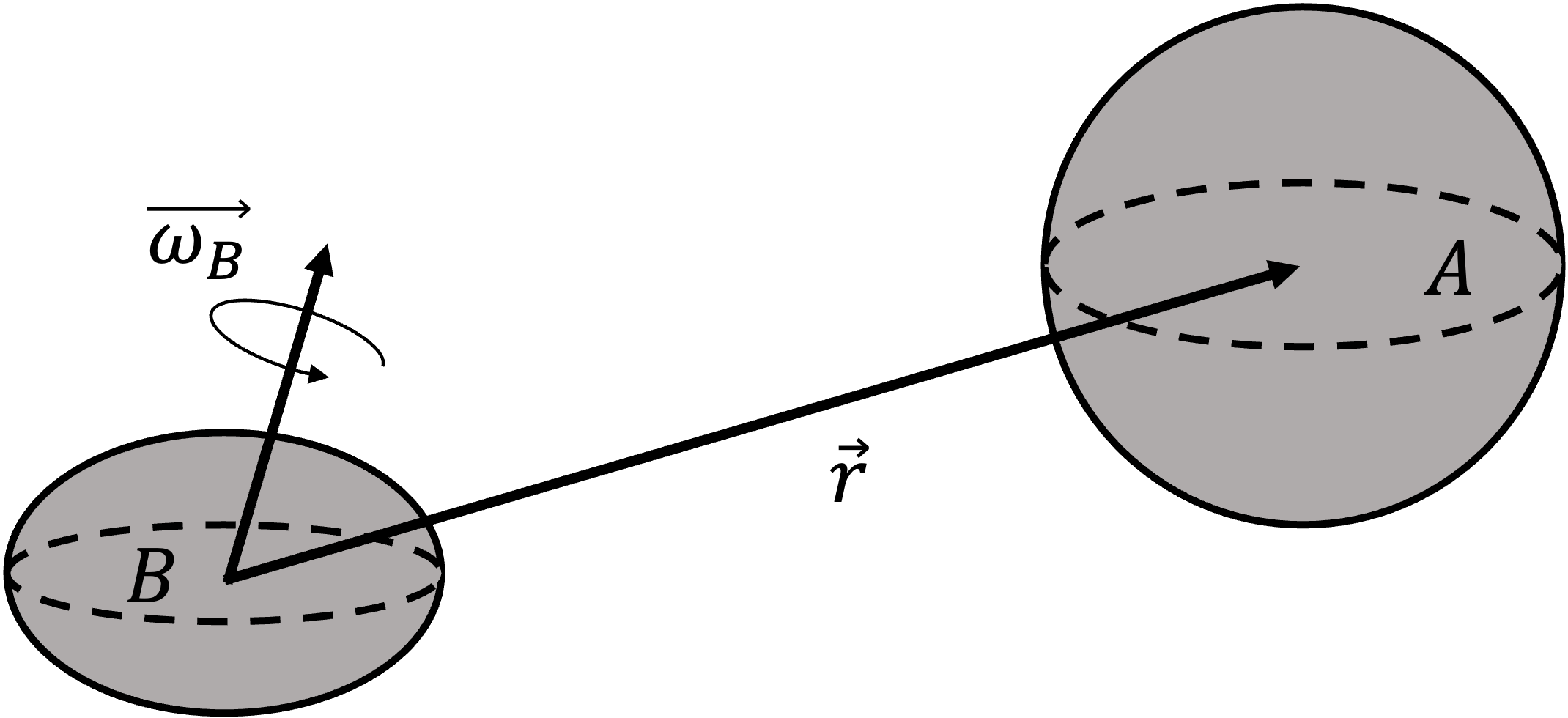} 
   \caption{Diagram showing the dynamic model.}
   \label{model}
\end{figure}

%###################################################%###################################################
The mutual dynamics of binary asteroids are characterized by the F2BP, in which the orbit and attitude of the bodies are coupled. This leads to complex dynamics, and various models have been developed to simulate these systems with varying trade-offs between fidelity and computational cost. Since we are concerned with timespans of many years, it is necessary to select a more basic model at the cost of reduced fidelity. In this context, we are more concerned with the system's qualitative behavior over long time periods rather than short-term accuracy, so this is a fine compromise. As such, we model Didymos as a spherical primary and Dimorphos as an ellipsoidal secondary, which allows for full 3D dynamics with an elongated secondary without becoming too computationally expensive. We will validate this model against a high-fidelity model in Section \ref{sec:gubas}. Fig. \ref{model} shows a diagram of the system, where body $A$ is Didymos and body $B$ is Dimorphos. In later discussions, quantities with subscripts $A$ or $B$ specify those for the designated bodies.

The equations of motion are straightforward and we omit any derivation, instead referring the reader to \cite{scheeres2006relative}. These equations are defined in the body-fixed frame of the secondary. We have six degrees of freedom, being the relative separation and the rotation of Dimorphos. The equations of motions are
\begin{equation}
    \ddot{\vec{r}}+2\vec{\omega}_B\times\dot{\vec{r}}+\dot{\vec{\omega}}_B\times\vec{r}+\vec{\omega}_B\times(\vec{\omega}_B\times\vec{r})=\mathcal{G}(M_A+M_B)\frac{\partial U}{\partial \vec{r}}
\end{equation}
\begin{equation}
    \mathbf{I}_B\cdot\dot{\vec{\omega}}_B+\vec{\omega}_B\times\mathbf{I}_B\cdot\vec{\omega}_B=-\mathcal{G}M_AM_B\vec{r}\times\frac{\partial U}{\partial \vec{r}}
\end{equation}
where $\mathbf{I}_B$ is the inertia tensor of the secondary, which is a simple diagonal matrix in the secondary's body-fixed frame. $U$ is the gravitational potential around the ellipsoidal Dimorphos, and to ease computation time we use a second degree expansion in the form of MacCullagh's formula \citep{murray1999solar}:
\begin{equation}
    U = -\frac{\mathcal{G}M_AM_B}{r}-\frac{\mathcal{G}M_A(A+B+C-3\Phi)}{2r^3}
\end{equation}
where $A$, $B$, and $C$ are respectively the minimum, intermediate, and maximum principal inertia values of Dimorphos, and $\Phi$ is a quantity defined by
\begin{equation}
    \Phi = \frac{Ax^2+By^2+Cz^2}{r^2}
\end{equation}
with $(x,y,z)$ being the Cartesian coordinates of the primary in the secondary's body-fixed frame so that $\vec{r} = x\hat{i}+y\hat{j}+z\hat{k}$.

We next need to introduce the methods of dissipation through non-rigid processes, both through tidal torque and NPA rotation, which rise from a combination of deformation, rotation, and translation of the bodies \citep{hirabayashi2022dynamics}. We ignore any surface motion on both the primary and secondary, which includes rotation-induced granular motion on the secondary's surface and any associated body reshaping, which changes the gravitational potential energy \citep{agrusa2022dynamical,agrusa2022rotation}, tidal saltation and YORP-induced landslides on the primary \citep{harris2009shapes}, and boulder movement on either body \citep{brack2019modeling}, which would also dissipate energy. We will assume any reshaping and surface motion to be small and intermittent, and the energy dissipated by these events to be negligible over time. So our estimates on damping times for Didymos can be considered conservative for a given set of material properties since additional mechanisms will only increase the dissipation rate.

\subsection{Tidal Torque} %###################################################
To describe energy dissipation from the system, we introduce equations for tidal torque to add to our dynamic model. In selecting a basic model for tidal torque, we have two choices: the constant $Q$ model, in which the rate of dissipation is driven by the ratio of tidal quality factor $Q$ and the simple Love number $k_2$ \citep{murray1999solar}, or the constant time lag model, in which the angle between the tidal bulge and the line connecting the two bodies is a constant $\Delta t$ \citep{mignard1979evolution,hut1981tidal}. Based on the physics of our problem setup, the secondary will librate about the synchronous configuration, and thus a constant lag angle would be inappropriate, as the lag angle should oscillate as a result of the libration. For this reason we select the constant $Q$ model, which is the same model adopted by \cite{jacobson2011dynamics}, in which the tidal torque is defined as:
\begin{equation}
    \Gamma = -\text{sign}(\omega-\omega_{orb})\frac{3}{2}\bigg(\frac{3}{4\pi\rho}\bigg)^2\frac{GM_A^2M_B^2}{r^6R}\frac{k}{Q}
	\label{torque}
\end{equation}
where the body's angular velocity is $\omega$, the orbit's angular rate is $\omega_{orb}$, $R$ is the reference radius for the body, $\rho$ is its density, and $Q/k$ is the tidal dissipation ratio. The tidal quality factor $Q$ is related to the tidal bulge lag angle ($Q\sim1/\sin{\epsilon}$), while the love number $k_2$ describes the level of body deformation due to the tidal potential. Henceforth we drop the subscript 2 on the Love number for simplicity. A large value of $Q/k$ corresponds to a more stiff body that dissipates more slowly. In reality, the tidal dissipation is far more complicated than simply selecting constant values for the unknown $Q/k$ values. As pointed out by \cite{efroimsky2015tidal}, tidal dissipation in binary asteroids, including rubble piles, may be governed primarily by the body's viscosity, rather than rigidity. Others, including \cite{goldreich2009tidal} and \cite{nimmo2019tidal}, argue that friction is a critical parameter. Since there is no current estimate for the viscosity of rubble pile asteroids to the authors' knowledge, we adopt the friction approach. Furthermore, while many studies assume the quality number $Q$ to be constant, this parameter depends on the tidal frequency. Further complicating this relationship is the fact that the tidal quality number is not a linear function of the tidal frequency, and can either increase or decrease with the frequency \citep{ferraz2013tidal}. We also are left with the problem of the tidal lag angle oscillation. To address this we adopt the same solution as \cite{jacobson2011dynamics}; we will linearize the tidal torque around the point where $(\omega-\omega_{orb})$ is near zero so the torque does not immediately switch signs as the secondary librates (see Appendix C therein for details on this linearization). This linearization is necessary, as the tidal bulge is a physical phenomenon and requires a finite time to cross between leading or trailing the tide-raising body. 

Note that \cite{taylor2010tidal} point out the simple tidal model assumes the two bodies are widely separated, whereas the separation between Didymos and Dimorphos is only slightly larger than 3 primary radii. In their work, \cite{taylor2010tidal} calculate that higher order terms in the tidal potential speed up the process of tidal evolution. However, they also find that uncertainties in the system, particularly surrounding $Q/k$, dominate over the higher order tidal expansion. Thus, we continue with the simple tidal model given the large uncertainty associated with the bulk system density and physical properties, while keeping in mind higher order terms in the tidal model will only increase the rate of damping in the system. Thus, the error associated with this tidal model is considered to be secondary to the considerable uncertainty on the $Q/k$ coefficient for our purposes.

Unfortunately, given the lack of knowledge on the physical parameters of rubble piles, particularly their viscosity, we cannot calculate an accurate value for $Q/k$. For lack of a better option, we surrender ourselves to the typical simplifications surrounding the factor $Q/k$, and we turn to the work by \cite{nimmo2019tidal}, who derive an estimate for a constant $Q/k$ for rubble pile binary asteroids, which can be approximated by
\begin{equation}
    \frac{Q}{k}\approx300R
\end{equation}
for $R$ in meters. This leads to a value for the primary $Q_A/k_A \approx 1\times10^5$ and for the secondary $Q_B/k_B \approx 2.5\times10^4$. Note this expression for $Q$ is frequency dependent and derived for a non-synchronous binary system. However, we again emphasize there is large uncertainty associated with these values so this definition serves as a first-order approximation, as the error from uncertainties dominates over the error from the assumptions. Furthermore, these values are consistent with existing estimates for small bodies in the literature \citep{brasser2020efficient, jacobson2011long, scheirich2015binary, scheirich2021satellite}. However, there is not a consensus on this linear scaling. For example, \cite{goldreich2009tidal} propose an inverse scaling of $Q/k$ with $R$, and even in their own work \cite{nimmo2019tidal} point out a scaling with $R^{3/2}$ may be more accurate. Another consideration is if Dimorphos turns out to be monolithic instead of a rubble pile, its $Q/k$ value would likely be orders of magnitude higher \citep{goldreich2009tidal}. Hence, the large uncertainty in $Q/k$ dominates over other errors associated with our model, and it is futile to develop a high fidelity tidal model while limited by this unknown parameter. Given the large uncertainty and lack of consensus around $Q/k$, we adopt the linear scaling only as nominal parameters, and subsequently investigate how varying $Q/k$ for both the primary and secondary affects the system behavior later in Section \ref{sec:QK}.

Returning to the tidal torque equation, this model is still only defined in 2-dimensions and we wish to extend this to a full 3-dimensional analysis, as out-of-plane rotation of the secondary is a possibility. This can be done with only a few corrections to the classic model. To start, we need to define a vector for the torque direction. The torque will act to push the spin rates of the asteroids into the synchronous equilibrium, but physically cannot act in the direction of the position vector of the secondary relative to the primary. We define the relative spin rate of a body:
\begin{equation}
	\dot{\vec{\phi}} = \vec{\omega}-\vec{\omega}_{orb}.
\end{equation}
We can then use this spin vector as the vector along which the torque acts, with a small correction so the torque in the radial direction is zero:
\begin{equation}
	\hat{\Gamma} = -\frac{\dot{\vec{\phi}} - (\dot{\vec{\phi}}\cdot\hat{r})\hat{r}}{|\dot{\vec{\phi}} - (\dot{\vec{\phi}}\cdot\hat{r})\hat{r}|}
	\label{gammahat}
\end{equation}

This formulation also takes care of the sign of the torque, as the torque will generally act in the direction opposite the relative spin rate, so that a secondary rotating faster than the orbit rate is slowed, while a secondary rotating slower will be sped up. Furthermore, any out-of-plane rotation is countered by the torque, with the exception of any spin about the relative position vector, as tidal torque can only act perpendicular to this direction. Thus, we don't need any further consideration on the sign of the torque and can remove the $-\text{sign}(\omega-\omega_{orb})$ expression from Eq. \ref{torque} and substitute Eq. \ref{gammahat} in its place.

While we formulated the 3D torque expression with the secondary in mind, it is also equally applicable to the primary, as tidal dissipation will ultimately drive the system to the doubly synchronous state. So we have developed expressions for the torque on both bodies. However, to accurately express the equations of motion we also need to consider the torque on the orbit. By the conservation of angular momentum, the torque on the orbit is simply
\begin{equation}
\vec{\Gamma}_{orb} = -(\vec{\Gamma}_A+\vec{\Gamma}_B).
\end{equation}
However, to include this in the orbital equation of motion we will need to calculate this torque's effect on $\ddot{\vec{r}}$. Turning to the orbital angular momentum we know
\begin{equation}
\dot{\vec{H}} = \frac{d}{d t}(m\vec{r}\times\dot{\vec{r}}) = \vec{\Gamma}_{orb}
\end{equation}
where $m = \frac{M_AM_B}{M_A+M_B}$. This gives
\begin{equation}
m\vec{r}\times\ddot{\vec{r}} = \vec{\Gamma}_{orb}.
\end{equation}
We would like to solve this equation for $\ddot{\vec{r}}$ to update the orbital equation of motion with the tidal torque. We note that
\begin{equation}
\vec{r}\times(\vec{r}\times\ddot{\vec{r}}) = (\vec{r}\cdot\ddot{\vec{r}})\vec{r} - (\vec{r}\cdot\vec{r})\ddot{\vec{r}} = \vec{r}\times\frac{\vec{\Gamma}_{orb}}{m}
\end{equation}
which can be arranged to find:
\begin{equation}
\ddot{\vec{r}} = \frac{\vec{\Gamma}_{orb}\times\vec{r}}{mr^2} + \frac{\vec{r}\cdot\ddot{\vec{r}}}{r^2}\vec{r}.
\end{equation}
However, this is not a unique solution, as for any real number $K$ we can find
\begin{equation}
\ddot{\vec{r}} = \frac{\vec{\Gamma}_{orb}\times\vec{r}}{mr^2} + K\frac{\vec{r}\cdot\ddot{\vec{r}}}{r^2}\vec{r}
\end{equation}
 is also a valid solution as a result of the dot product. Thus, we must solve the problem of a non-unique solution. We note that whenever the torque $\Gamma_{orb}$ is equal to zero, there should be no contribution to the equation of motion. So we choose $K=0$ to enforce this constraint (alternatively, we force the acceleration to be perpendicular to the position vector, so that $\vec{r}\cdot\ddot{\vec{r}}=0$), and we are left with
\begin{equation}
\ddot{\vec{r}} = \frac{\vec{\Gamma}_{orb}\times\vec{r}}{mr^2} = \vec{\gamma}_{orb}.
\end{equation}
where we introduce the term $\vec{\gamma}_{orb}$ as the acceleration due to the torque on the orbit. This makes sense given the physical context, where the tides should not have any effect in the radial direction. Note that this formula must be calculated in an inertial frame as it is derived from the orbit angular momentum. Once computed in an inertial frame, it can be transformed into the secondary body-fixed frame for use in the equations of motion.

\subsection{NPA Rotation} %###################################################
Any NPA rotation of the secondary will act to dissipate energy while keeping its angular momentum constant, until its maximum principal inertia axis is aligned with its angular momentum. There is a wide variety of work that has considered this problem \citep{burns1973asteroid, molina2003energy, pravec2005tumbling, ershkov2019dynamics}, but for the most applicable to this work we turn to \cite{breiter2012stress}, who calculate the rate of energy dissipation in a triaxial ellipsoid rotating in either long-axis mode or short-axis mode. The full expression for energy dissipation in a triaxial ellipsoid undergoing NPA rotation is quite complicated, and we report a condensed version here:
\begin{equation}
    \dot{E}_{NPA} = \frac{a^4\rho M_B\tilde{\omega}_B^5}{\mu Q}\Psi
\end{equation}
where $a$ is the secondary's longest semiaxis, $\tilde{\omega}_B$ is its nominal rotation rate, and $\Psi$ is a complicated function of the secondary's shape, Poisson's ratio, and an elliptic modulus. We omit the details here and refer the reader to \cite{breiter2012stress}, for the full expression. This expression uses the Lamé constant $\mu$ instead of the Love number $k$, but we can relate our $Q/k$ to $\mu Q$ through \citep{murray1999solar, nimmo2019tidal}:
\begin{equation}
    \mu Q \sim \mathcal{G}\frac{Q}{k} \rho^2 R^2.
\end{equation}
At the risk of sounding repetitive, we again highlight the shortfall of this relationship as pointed out by \cite{efroimsky2015tidal}, as $Q$ is frequency dependent. Thus, the $Q$ selected for the tidal dissipation model is not necessarily the same $Q$ for NPA rotation, again introducing considerable uncertainty. However, we assume the driving frequencies for tidal and NPA dissipation are the same, and again for lack of a better option we adopt the same $Q$ for both methods of dissipation.

Since the equations of motion are defined in the secondary's body-fixed frame, the energy dissipation due to NPA rotation will apply an `effective torque' to rotate the secondary's angular momentum to align it with its maximum principal inertia axis while keeping the magnitude constant.

To obtain the effective torque, we start with the energy dissipation equation
\begin{equation}
	\dot{E}=\vec{\omega}_B\cdot I_B \dot{\vec{\omega}}_B.
\end{equation}
Next, we expand the equation $\dot{\vec{H}}_B=\vec{\Gamma}_{NPA}$ for secondary angular momentum $H_B$:
\begin{equation}
\vec{\Gamma}_{NPA} = I_B\dot{\vec{\omega}}_B + \vec{\omega}_B\times I_B\vec{\omega}.
\end{equation}
By taking the dot product of this equation with $\vec{\omega}_B$, we can find an expression for the effective torque as a function of $\dot{E}$:
\begin{equation}
	\vec{\omega}_B\cdot\vec{\Gamma}_{NPA}=\dot{E}_{NPA}.
\end{equation}
Of course, due to the dot product, this is not a complete expression as we still need to define the direction of the effective torque $\vec{\Gamma}_{NPA}$. Since NPA dissipation will not change the angular momentum magnitude, only the angular momentum's direction relative to the body-fixed axes of the secondary, the effective torque must be perpendicular to the angular momentum. As a conceptual illustration, imagine a satellite undergoing torque-free tumbling. In this scenario, the satellite's angular momentum vector is constant, while dissipation reorients the satellite so that its principal axis is aligned with the angular momentum (and its kinetic energy is minimized). In the same sense, within the body-fixed frame of the secondary, its angular momentum will appear to rotate until its aligned with the principal axis, while maintaining a constant magnitude. This is accomplished by applying an effective torque perpendicular to the angular momentum, such that this torque only rotates the angular momentum vector, but does not scale it. We define an intermediate vector
\begin{equation}
	\vec{\kappa} = \vec{H}_B\times\hat{z}
\end{equation}
where $\hat{z}$ is the secondary's maximum principal inertia axis. We then use this intermediate vector to define the torque unit vector
\begin{equation}
	\hat{\Gamma} = \frac{\vec{H}_B\times\vec{\kappa}}{|\vec{H}_B\times\vec{\kappa}|}.
\end{equation}
The effective torque will act to rotate the angular momentum vector $\vec{H}_B$ until it is aligned with the secondary's $\hat{z}$ direction. The effective torque is then calculated as
\begin{equation}
	\vec{\Gamma}_{NPA} = \frac{\dot{E}}{\vec{\omega}_B\cdot\hat{\Gamma}}\hat{\Gamma}.
\end{equation}
Note that in reality, the secondary angular momentum vector will not rotate and instead the body-fixed coordinate axes of the secondary will rotate to align $\hat{z}$ with the angular momentum vector. However, if we define our equations of motion in the secondary body-fixed frame as we have done, this effective torque will be appropriate to use.

\subsection{Comparison} %###################################################
\begin{figure}[ht!]
   \centering
   \includegraphics[width = 3in]{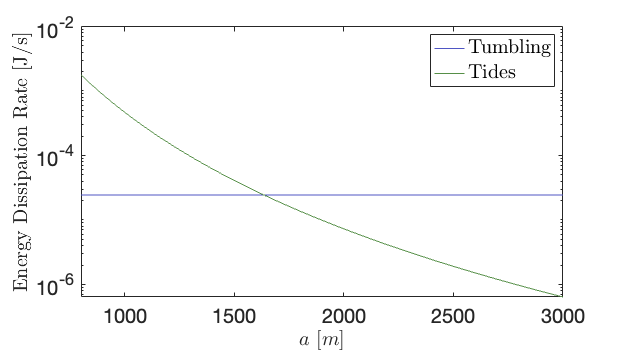} 
   \caption{The rate of energy dissipation through tidal torque compared to NPA rotation as a function of separation. Tidal torque has orders of magnitude more of an effect than NPA rotation for a close binary system like Didymos.}
   \label{dissipation_compare}
\end{figure}

In order to compare our two methods of energy dissipation, we plot the rate of energy dissipation in the secondary as a function of the separation distance for a body rotating with a spin axis offset from its major principal axis by approximately $32^\circ$. This is near the switching point between long-axis and short-axis mode rotation, and thus NPA dissipation is near a maximum here. Fig. \ref{dissipation_compare} shows the energy dissipation rates for the two mechanisms, highlighting that tidal dissipation is generally more than an order of magnitude stronger than NPA rotation for a close binary like Didymos. Despite this contrast, we still include the effects of NPA rotation as it does not significantly slow down the integration wall time.

While close systems like Didymos are dominated by tidal torque, wider systems see near-equal contributions from both tidal torque and NPA rotation, and NPA dissipation becomes dominant at very wide separations. The secondary shape and spin-axis obliquity will affect the rate of NPA rotation energy decay, but for Didymos the tidal torque will always be stronger owing to the close proximity of its secondary. 

\begin{figure}[ht!]
   \centering
   \includegraphics[width = 3in]{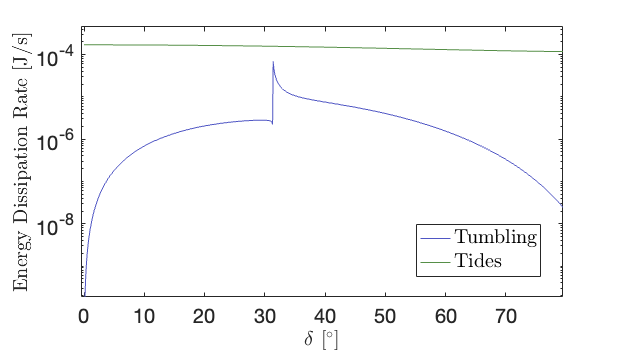} 
   \caption{The rate of energy dissipation through tidal torque compared to NPA rotation as a function of secondary spin offset angle. The NPA dissipation is greatest near the discontinuity, where the secondary is switching between long-axis and short-axis rotation modes.}
   \label{dissipation_compare_oblq}
\end{figure}

Looking at a system with a semimajor axis of 1200 m, we next examine how the secondary's spin offset angle affects dissipation. Fig. \ref{dissipation_compare_oblq} shows the dissipation rate as a function of the angle between the secondary's spin axis and its principal moment of inertia axis (we call this angle $\delta$, and a system with $\delta=0^\circ$ is in principal axis spin). We see tidal dissipation is largely independent of the spin axis offset, but NPA rotation depends strongly on this angle. There is a discontinuity where the secondary switches from short-axis to long-axis mode rotation. Dissipation then quickly drops to zero as spin approaches either perfectly major ($\delta=0^\circ$) or minor ($\delta=90^\circ$) axis rotation.

\subsection{Equations of Motion} \label{sec:EOM} %###################################################
We can now add our dissipation mechanisms to our equations of motion, which will model the energy dissipation due to tides and NPA rotation while keeping the system's angular momentum constant:
\begin{equation}
\begin{split}
	\ddot{\vec{r}}+2\vec{\omega}_B\times\dot{\vec{r}}+\dot{\vec{\omega}}_B\times\vec{r}+\vec{\omega}_B\times(\vec{\omega}_B\times\vec{r})\\=\mathcal{G}(M_A+M_B)\frac{\partial U}{\partial \vec{r}}+\vec{\gamma}_{orb}
\end{split}
\end{equation}
\begin{equation}
	\mathbf{I}_B\cdot\dot{\vec{\omega}}_B+\vec{\omega}_B\times \mathbf{I}_B \cdot\vec{\omega}_B=-\mathcal{G}M_AM_B\vec{r}\times\frac{\partial U}{\partial \vec{r}}+\vec{\Gamma}_B+\vec{\Gamma}_{NPA}.
\end{equation}
The uncoupled rotational dynamics equation of the spherical primary is simply
\begin{equation}
	\mathbf{I}_A\dot{\vec{\omega}}_A = \vec{\Gamma}_A,
\end{equation}
which is necessary to include to enforce the conservation of angular momentum.

\begin{figure}[ht!]
   \centering
   \includegraphics[width = 3in]{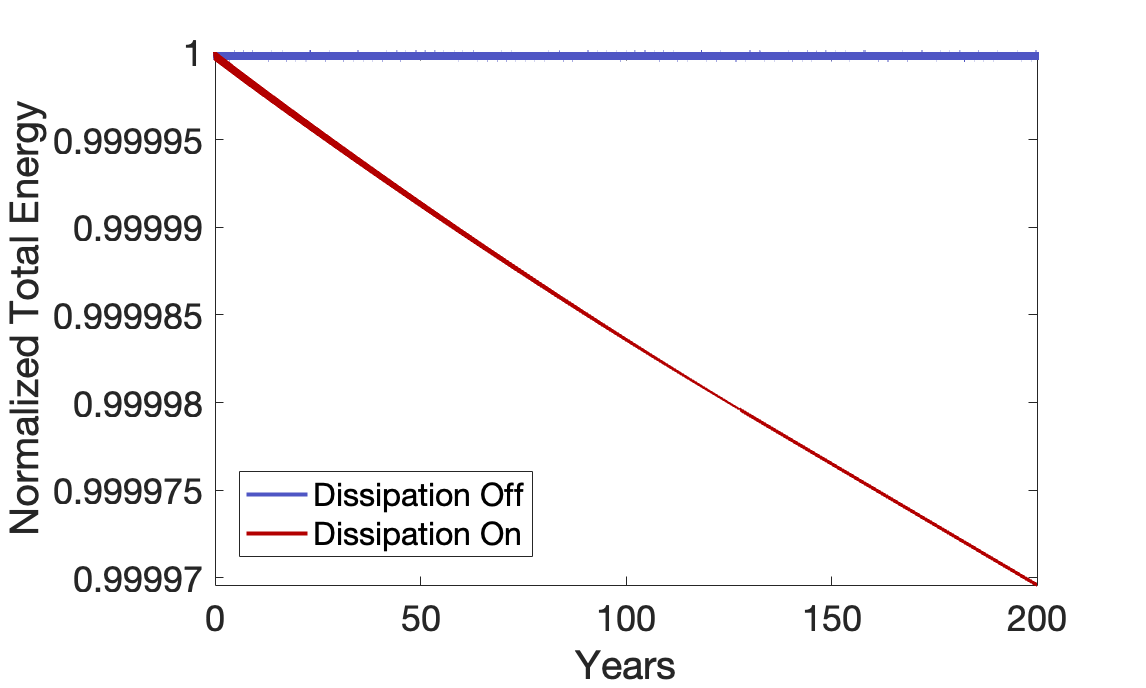} 
   \caption{The total system energy, normalized by the initial energy, over 200 years comparing the dissipative to the non-dissipative case. This demonstrates the energy dissipation is a real effect and not caused by numerical error over this long time span.}
   \label{integrator_check}
\end{figure}

For the sphere-ellipsoid model, we use a variable-step, variable-order Adams-Bashforth-Moulton predictor-corrector integrator to propagate the equations of motion. This integrator conserves the system energy (in the non-dissipative case) to within $5\times10^{-5}\%$ over 200 years. As an illustration, Fig. \ref{integrator_check} shows the total system energy calculated by propagating the equations of motion using this integrator, for both the dissipative and non-dissipative cases. This demonstrates that the integrator conserves energy in the non-dissipative case and is accurate enough to capture the secular effects caused by energy dissipation.

\section{Energy Dissipation} \label{sec:Dissipation} %###################################################
%###################################################%###################################################
In investigating the energy dissipation of a librating binary asteroid, we must consider two dynamical regimes: stable libration or NPA rotation. A system in the stable regime will only see fluctuations in secondary rotation about its principal axis, whereas an unstable system will experience rotation about all three axes. For a uniform bulk density, the dynamical regime of the system depends on the shape of the secondary, as illustrated in Fig. \ref{beta3_impact}. We will select a system from both the stable and unstable region to carry out long-term simulations to investigate how a system dissipates its libration and returns to a synchronous equilibrium. For the stable system, we select a secondary with axis ratios $a/b=1.2$, $b/c=1.1$, and for the unstable system we choose $a/b=1.4$, $b/c=1.3$. Both of these shapes are within the predicted values for Dimorphos \citep{pravec2022photometric}. Note that energy dissipation in the primary is generally uniform across all shapes, so we thus exclude it from our analysis for simplicity.

\subsection{Stable System} %###################################################
\begin{figure}[ht!]
   \centering
   \includegraphics[width = 3in]{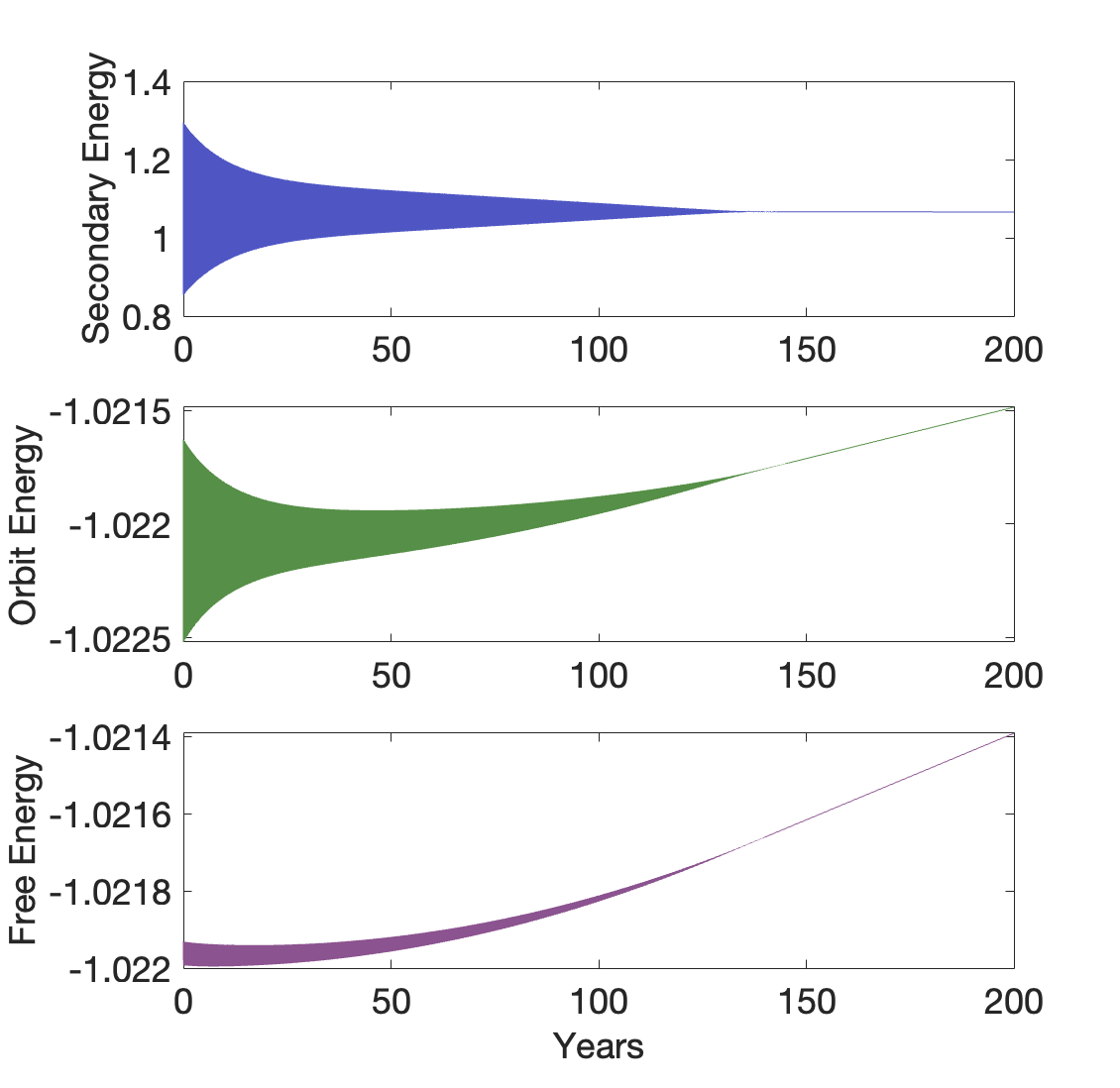} 
   \caption{The secondary rotational energy (top), orbit energy (middle), and free energy (bottom) of the stable system with secondary $a/b=1.2$, $b/c=1.1$. The energies are normalized by the respective pre-impact equilibrium values. The oscillations within the energy are damped at the same rate across the secondary, orbit, and free energies. Here $Q_A/k_A \approx 1\times10^5$ and $Q_B/k_B \approx 2.5\times10^4$.}
   \label{energy_stable}
\end{figure}

For the stable system, we simulate the dynamics for 200 years ($\sim150,000$ orbit periods), during which the system returns to an equilibrium and begins to secularly evolve. We first examine the system energy, shown in Fig. \ref{energy_stable}. We plot the secondary rotational energy, the orbit energy, and the free energy. We define the free energy simply as the sum of the secondary rotational and orbit energies, or alternatively as the total energy sans the primary rotational energy. Since the primary rotational energy will uniformly decrease as a result of tides, the free energy is a better metric to examine the system's evolution. We see oscillations in the energy after the perturbation, which damp out as the system settles into a new equilibrium. Interestingly, the free energy decreases for a time before increasing. This indicates that dissipation acts to equilibrate the system faster than secularly evolve it. This is apparent in the exponential decrease in the secondary and orbit energy oscillations, as these values converge to a mean faster than the mean itself evolves.

\begin{figure}[ht!]
   \centering
   \includegraphics[width = 3in]{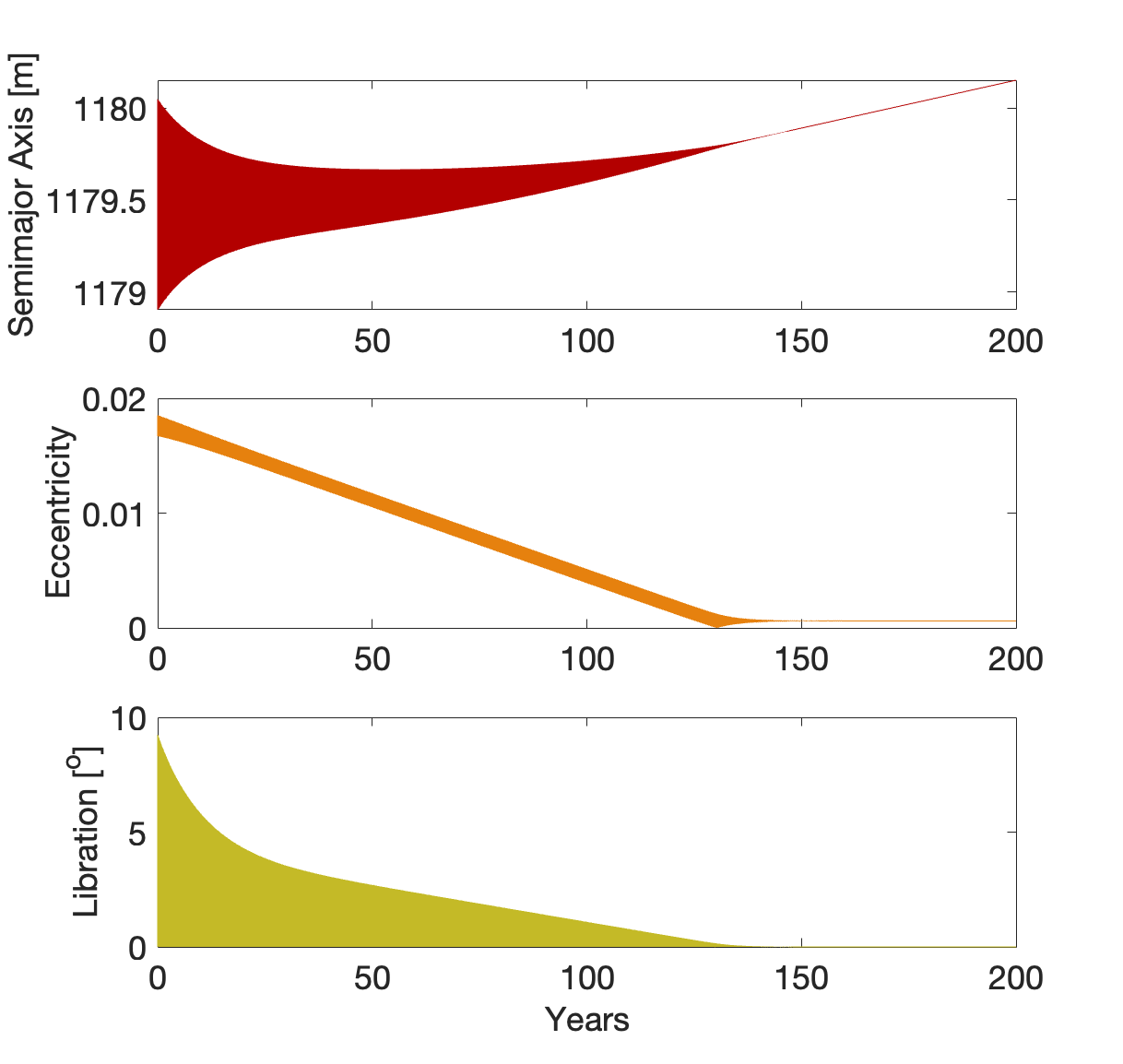} 
   \caption{The Keplerian semimajor axis (top), eccentricity (middle), and libration angle (bottom) of the stable system with secondary $a/b=1.2$, $b/c=1.1$. The oscillations in semimajor axis and libration are damped at the same rate the eccentricity approaches its non-zero equilibrium value due to spin-orbit coupling. Here $Q_A/k_A \approx 1\times10^5$ and $Q_B/k_B \approx 2.5\times10^4$.}
   \label{elements_stable}
\end{figure}

This behavior is also apparent in Fig. \ref{elements_stable}, where we plot the Keplerian semimajor axis and eccentricity, along with the secondary libration angle. Again, we see the oscillations in the semimajor axis exponentially damping before any secular evolution is apparent. These oscillations damp at the same rate the eccentricity approaches its equilibrium value (recall that due to spin-orbit coupling, the equilibrium eccentricity is small but non-zero). This is also the same rate the libration damps to zero. In the libration angle we again see an initial exponential decrease. 

As a result of spin-orbit coupling and the unique dynamics of binary asteroids, we see a new dynamical regime not previously studied in tidal analyses. It appears dissipation first acts to drive the system back toward an equilibrium by damping the eccentricity and oscillations in the system caused by libration, before the more classical secular tidal behavior is seen. Thus, rather than eccentricity and semimajor axis evolving concurrently as in classical tidal theory (e.g. \cite{goldreich2009tidal}), the eccentricity is damped to a minimum before the semimajor axis evolves.

\subsection{Unstable System} %###################################################
\begin{figure}[ht!]
   \centering
   \includegraphics[width = 3in]{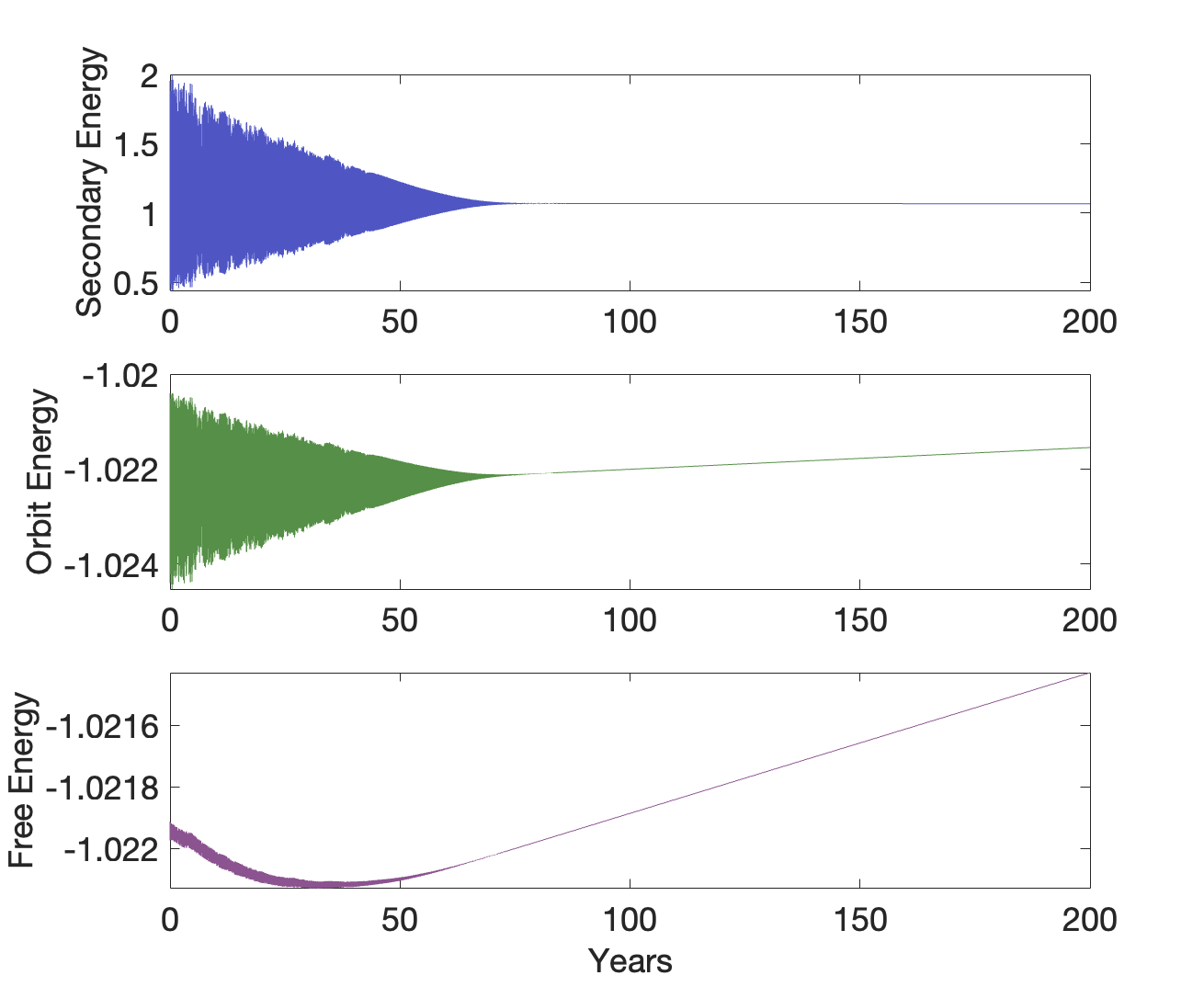} 
   \caption{The secondary rotational energy (top), orbit energy (middle), and free energy (bottom) of the unstable system with secondary $a/b=1.4$, $b/c=1.3$. The energies are normalized by the respective pre-impact equilibrium values. The oscillations within the energy are damped at the same rate across the secondary, orbit, and free energies. Here $Q_A/k_A \approx 1\times10^5$ and $Q_B/k_B \approx 2.5\times10^4$.}
   \label{energy_unstable}
\end{figure}

We next perform the same simulation for the unstable system. The system energy is shown in Fig. \ref{energy_unstable}, where again we plot the secondary rotational energy, the orbit energy, and the free energy. Interestingly, on a comparable time scale as the stable system, the unstable system also returns to an equilibrium configuration. A notable difference is the oscillations in energy appear to converge linearly, unlike the stable system which experienced an exponential decay of oscillations. In the unstable system we again see an initial decrease in the free energy before it begins to increase as the orbit expands. 

\begin{figure}[ht!]
   \centering
   \includegraphics[width = 3in]{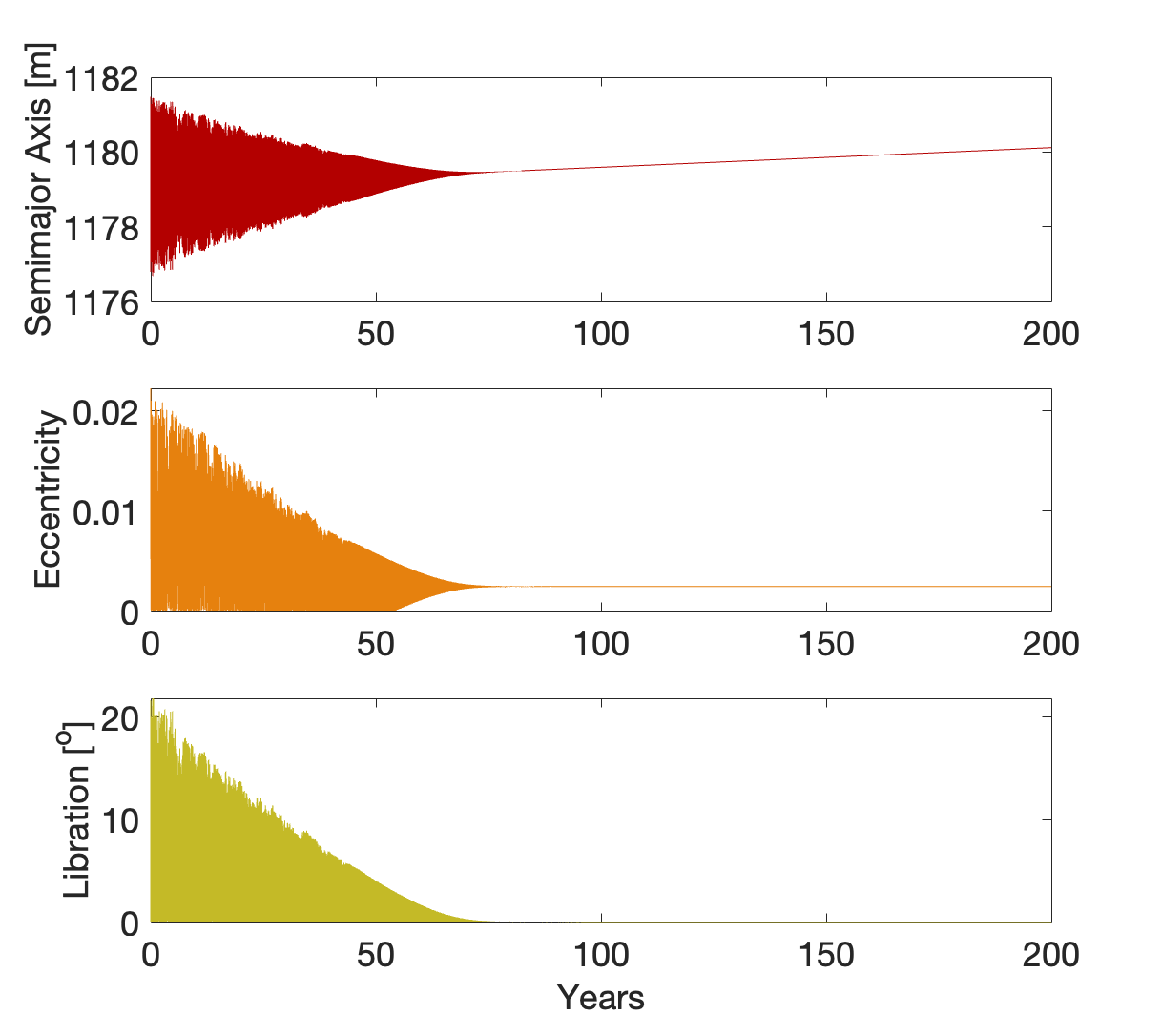} 
   \caption{The semimajor axis (top), eccentricity (middle), and libration angle (bottom) of the unstable system with secondary $a/b=1.4$, $b/c=1.3$. The oscillations in semimajor axis and libration are damped at the same rate the eccentricity approaches its equilibrium value due to spin-orbit coupling. Here $Q_A/k_A \approx 1\times10^5$ and $Q_B/k_B \approx 2.5\times10^4$.}
   \label{elements_unstable}
\end{figure}

In Fig. \ref{elements_unstable}, we plot the semimajor axis, eccentricity, and libration angle of the unstable system. Again we see a linear, rather than exponential, decay in these elements. Once more, the timescale of damping is approximately uniform across all these elements, also consistent with the energy envelope decay times, highlighting the strength of the spin-orbit coupling. The binary asteroid cannot settle into an equilibrium while the libration amplitude is nonzero, or equivalently the eccentricity is not equal to its equilibrium value. This is consistent with the findings of \cite{meyer2021libration}, who point out the relationship between the secondary spin and the orbit oscillations.

An interesting takeaway from this result is both the stable and unstable systems reequilibrate on comparable timescales. While previous analyses have shown unstable rotation will slow dissipation \citep{wisdom1984chaotic,quillen2022non,naidu2015near}, that applies to non-synchronous rotation, whereas our system never leaves an on-average synchronous configuration. \cite{quillen2020excitation} predict that tumbling within the synchronous state does not reduce energy dissipation and in fact can enhance it, and our results are consistent with that finding. However, a unique finding in our analysis is the close relationship between the eccentricity, libration, and oscillation within the semimajor axis. This is due to the spin-orbit coupling in binary asteroids, as any deviation from an equilibrium spin state of the secondary will also affect the orbit.

\subsection{Analytic Models} %###################################################
\begin{figure}[ht!]
   \centering
   \includegraphics[width = 3in]{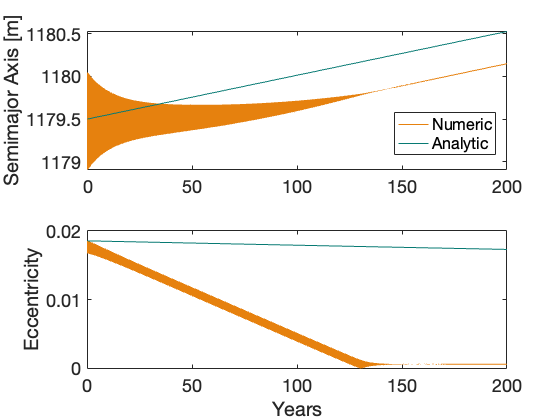} 
   \caption{The semimajor axis (top) and eccentricity (bottom) for the stable system ($a/b=1.2$, $b/c=1.1$), comparing the numeric results to an analytic model. While the secular trend of semimajor axis is consistent between the models, the analytic model does a poor job of describing the eccentricity evolution.}
   \label{analytic_stable_orbit}
\end{figure}

\begin{figure}[ht!]
   \centering
   \includegraphics[width = 3in]{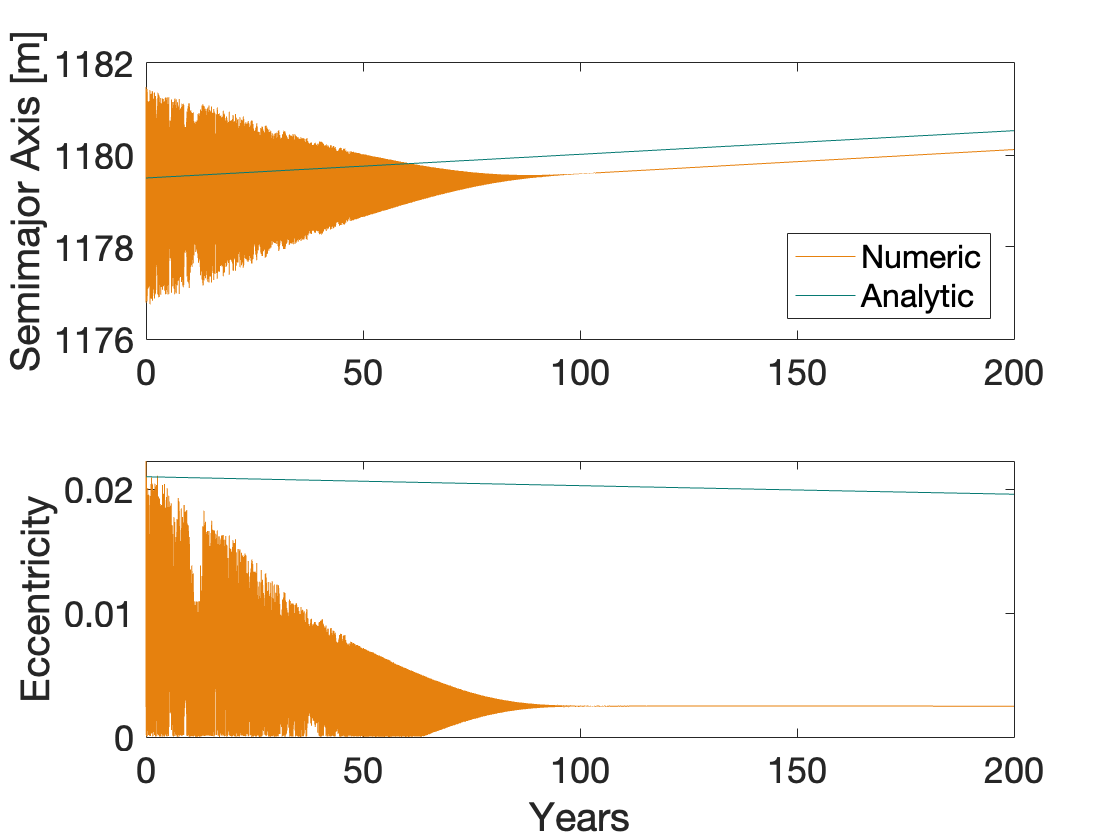} 
   \caption{The semimajor axis (top) and eccentricity (bottom) for the unstable system ($a/b=1.4$, $b/c=1.3$), comparing the numeric results to an analytic model. While the secular trend of semimajor axis is consistent between the models, the analytic model does a poor job of describing the eccentricity evolution.}
   \label{analytic_unstable_orbit}
\end{figure}

An important question is how this numeric model compares to more classical analytic models of energy dissipation. Here, we focus only on tidal dissipation since this is the dominant mechanism. Analytic equations for the evolution of semimajor axis and eccentricity of a binary system with $e\ll1$ undergoing tidal dissipation are reported in \cite{goldreich2009tidal}:

\begin{equation}
    \frac{\dot{a}}{a}=3\frac{k_A}{Q_A}\frac{M_B}{M_A}\bigg(\frac{R_A}{a}\bigg)^5n
    \label{analytic_sma}
\end{equation}

\begin{equation}
    \frac{\dot{e}}{e}=\frac{57}{8}\frac{k_A}{Q_A}\frac{M_B}{M_A}\bigg(\frac{R_A}{a}\bigg)^5n-\frac{21}{2}\frac{k_B}{Q_B}\frac{M_A}{M_B}\bigg(\frac{R_B}{a}\bigg)^5n
    \label{analytic_eccentricity}
\end{equation}
where $n$ is the mutual orbit mean angular velocity. Note in Eqs. \ref{analytic_sma} and \ref{analytic_eccentricity}, $a$ refers to the binary mutual orbit semimajor axis rather than the secondary's longest semiaxis as used elsewhere in this work. In Eq. \ref{analytic_eccentricity} for the eccentricity we see two terms: the first is the contribution from tides raised on the primary and the second is the contribution from tides raised on the secondary. The semimajor axis and eccentricity evolution is compared between the analytic model and our numeric results in Fig. \ref{analytic_stable_orbit} for the stable system and Fig. \ref{analytic_unstable_orbit} for the unstable system.

For both the stable and unstable systems, the analytic model does a good job describing the secular rate of change of semimajor axis, but fails to account for the initial oscillations in semimajor axis. The analytic model also fails to describe the eccentricity damping rate and predicts a much slower decrease in eccentricity than what we see in our numeric model. These results make sense in the context of this problem; in our numeric model, the primary matches the assumptions in the analytic model: a rapidly rotating sphere. However, our secondary does not match the assumptions as it is neither spherical nor exactly tidally locked. Thus, we would expect to match the semimajor axis drift as this is driven by tides on the primary, but not the eccentricity damping rate, which has contributions from tides on the secondary. Furthermore, as the semimajor axis oscillations are driven by libration of the secondary, the analytic model does not capture this behavior. Note that our results have a non-negligible eccentricity, so we expect some error in the analytic model.

The analytic model of \cite{goldreich2009tidal} makes no consideration of the secondary's libration. For this we turn to \cite{jacobson2013formation} who develop an expression for the damping rate of libration amplitude due to tides\footnote{The final result of this equation reported in \cite{jacobson2013formation} is missing the $1/\mathsf{C}$ term in the denominator.}:

\begin{equation}
    \dot{\Phi}_B = -\frac{k_B\omega_l\Phi_B^2}{QS\mathsf{C}\sin{2\Phi_B}}\bigg(\frac{R_A}{\tilde{a}}\bigg)^3
\end{equation}
where the libration frequency $\omega_l$ is defined as
\begin{equation}
    \omega_l=\frac{\pi\omega_d\sqrt{3S(1+s)}}{2\text{K}(\sin^2\Phi_B)}\bigg(\frac{R_A}{\tilde{a}}\bigg)^{3/2}.
\end{equation}
Here, $\omega_d=\sqrt{4\pi\rho G/3}$ is the spin disruption limit, $\text{K}(k^2)$ is the complete elliptic function of the first kind, $s=\mathsf{C}q^{2/3}(1+q)(R_A/\tilde{a})^2$ is the secondary perturbation term with mass fraction $q=m_B/m_A$. Here, $\tilde{a}$ is the binary mutual orbit semimajor axis in units of the primary radius. $S=(\mathsf{B}-\mathsf{A})/\mathsf{C}$ is a shape parameter where $\mathsf{A}<\mathsf{B}<\mathsf{C}$ are the dimensionless principal moments of inertia. For the derivation and in-depth discussion of this model we refer the reader to \cite{jacobson2013formation}.

\begin{figure}[ht!]
   \centering
   \includegraphics[width = 3in]{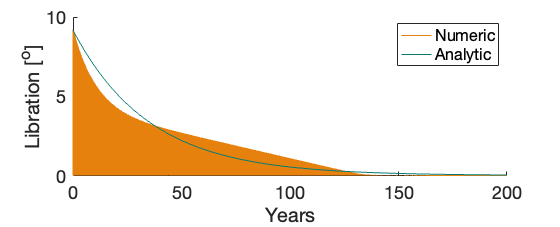} 
   \caption{The libration amplitude for the stable system ($a/b=1.2$, $b/c=1.1$), comparing the numeric results to an analytic model. We see good agreement between the models, and both converge to zero on the same timescale.}
   \label{analytic_stable_lib}
\end{figure}

\begin{figure}[ht!]
   \centering
   \includegraphics[width = 3in]{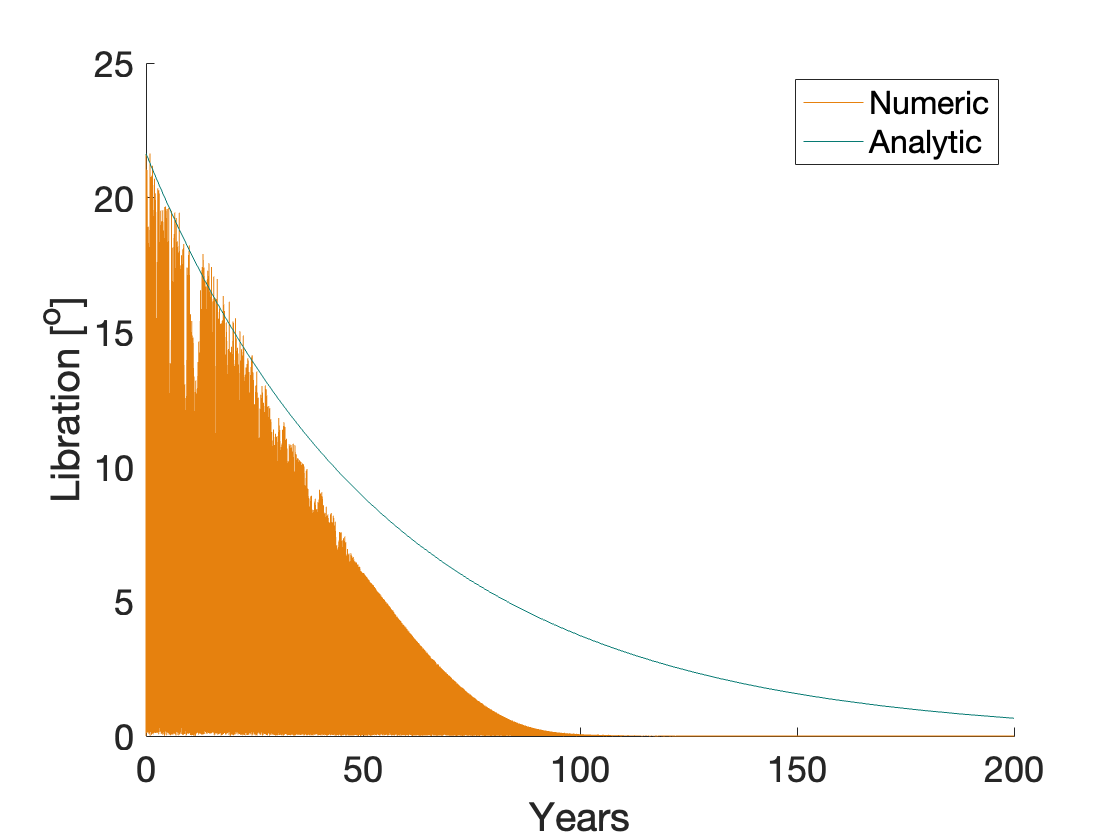} 
   \caption{The libration amplitude for the unstable system ($a/b=1.4$, $b/c=1.3$), comparing the numeric results to an analytic model. The numeric model, experiencing non-planar libration, dissipates to zero faster than the analytic model predicts for planar libration.}
   \label{analytic_unstable_lib}
\end{figure}

While this formula is derived for a planar, circular orbit, it can still accurately be applied to our stable system as this remains essentially planar with a small eccentricity. We see in Fig. \ref{analytic_stable_lib} that, for the stable system, there is good agreement between our numeric results and this analytic model. While the instantaneous dissipation rates can differ, the overall trend is similar and both models converge to zero in the same time frame. Applying this analytic model to the unstable system introduces error given the non-planar libration, but Fig. \ref{analytic_unstable_lib} still shows decent agreement. The biggest difference for the unstable system is the rate of dissipation, where the numeric model converges to zero faster than the analytic model. This indicates out-of-plane rotation of the secondary increases the rate of dissipation compared to purely planar libration, consistent with the findings of \cite{quillen2020excitation}.

While \cite{jacobson2013formation} use the same equations as \cite{goldreich2009tidal} for the orbit evolution (semimajor axis and eccentricity), their derivation of the libration damping rate sidesteps the problem with the eccentricity damping equation. Thus, the formula for libration damping gives a good approximation for the eccentricity damping since these quantities are closely related in the coupled problem. Of course, this formula is more accurate for the stable, planar system. But a decent analytic approximation of the system's evolution can be made using only the semimajor axis and libration amplitude equations.

We have so far only compared two secondary shapes. These two shapes have a similar damping timescale, but additional tests are required to determine if this is a coincidence or a more general property. Fig. \ref{other_shapes} in Appendix \ref{plots} compares the libration damping of six additional shapes of the secondary, and here we see across all 8 shapes tested the damping timescale is on the same order of magnitude regardless of the secondary's rotational stability.

\section{Validation} \label{sec:gubas} %###################################################
%###################################################%###################################################
While we have developed an efficient model for energy dissipation in two close rigid bodies, the sphere-ellipsoid model is simplistic and we must validate it against a higher-fidelity model. As a high-fidelity model, we use the General Use Binary Asteroid Simulator (\textsc{gubas}), which uses a fourth degree and order gravity field between two rigid bodies defined by polyhedral models \citep{davis2020doubly, davis2021gubas}. A fourth degree gravity expansion was previously found to accurately describe binary asteroid dynamics \citep{agrusa2020benchmarking}. While we use the polyhedral model for the Didymos primary, we continue using an ellipsoid secondary to test both stable ($a/b=1.2$, $b/c=1.1$) and unstable ($a/b=1.4$, $b/c=1.3$) systems. We have modified the \textsc{gubas} code to include the same tidal torque model as described in Section \ref{sec:dynamics}, but we do not include NPA rotation dissipation for efficiency as it is generally at least an order of magnitude smaller. We run these simulations for 5 years as this is generally long enough to compare trends, but longer simulations in \textsc{gubas} are prohibitively expensive to run.

In comparing these models, we use the set of 1-2-3 Euler angles (roll, pitch, yaw) to describe the secondary's orientation relative to the synchronous equilibrium. This allows for a more accurate comparison than using only the physical libration angle.

\subsection{Stable System}
For the stable system, we first examine the energy, where Fig. \ref{model_stable_energy} in Appendix \ref{plots} plots the secondary, orbit, and total energies. While, unsurprisingly, there are differences in the magnitudes of these quantities, their overall behavior is consistent. The rate of collapse of the secondary and orbit energies are similar between the models, and the secular trend of total energy is also consistent. There are larger oscillations in the total energy for the \textsc{gubas} model, but this is due to numerical noise, as the simplicity of the sphere-ellipsoid model allows us to use much tighter tolerances without a major sacrifice to computation cost.

We next examine the libration amplitude through 1-2-3 Euler angles (roll, pitch, yaw). Fig. \ref{model_stable_euler} in Appendix \ref{plots} plots the 1-2-3 Euler angles for both the sphere-ellipsoid and \textsc{gubas} models, but $\theta_1$ and $\theta_2$ remain nearly zero due to the stable configuration. Again, there are small differences in the libration magnitude, but the damping rate is nearly identical between these models. One notable difference is it appears $\theta_1$ and $\theta_2$, although nearly zero, only dissipate further in the sphere-ellipsoid model. However, we recall that the sphere-ellipsoid model uses a more stringent tolerance than \textsc{gubas} and we see less numeric noise (see Fig. \ref{model_stable_energy}). Thus, we expect when the out-of-plane angles are small enough as in this case, they will not further damp as a result of this numeric noise. Even in the sphere-ellipsoid model, the out-of-plane angles damp further but still do not reach exactly zero. The in-plane angle $\theta_3$, which has a significant amplitude, has a very strong agreement between the two models.

\subsection{Unstable System}
Next, we perform the same comparison for the unstable system. Fig. \ref{model_unstable_energy} in Appendix \ref{plots} plots the secondary, orbit, and total energy. Essentially, we see the same behavior as we did in the stable system, where there are clear, and expected, differences between the two models but the overall behaviors are very similar.

For the unstable system, where non-principal axis rotation is prevalent, we plot the 1-2-3 Euler angles of the secondary in Fig. \ref{model_unstable_euler} in Appendix \ref{plots}. Again we see close agreement between the sphere-ellipsoid and \textsc{gubas} models. Note the secondary rolls over occasionally in the \textsc{gubas} model and $\theta_1$ oscillates about $180^\circ$ instead of $0^\circ$, but this is only a small deviation from the behavior seen in the sphere-ellipsoid model.

While there are quantitative differences between the sphere-ellipsoid and \textsc{gubas} models, they share qualitatively the same behavior. This is expected, as the high-fidelity model simply includes additional perturbations to the dynamics compared to the simple sphere-ellipsoid model. Most importantly, we are mainly concerned with the overall trends caused by dissipation, which are consistent across the models.

\section{Short-Term Implications} \label{sec:Short Term} %###################################################
%###################################################%###################################################
More applicable to the AIDA collaboration is how much the Didymos system will evolve between the DART impact and the arrival of Hera to survey the system; this timespan is around 5 years. Thus, we next investigate how different secondary shapes cause different behaviors over this shorter timespan. Given the large number of systems we must test to cover possible secondary sizes, we return to the sphere-ellipsoid model for computational efficiency.

\subsection{Secondary Energy}
\begin{figure}[ht!]
  \centering
  \includegraphics[width = 3in]{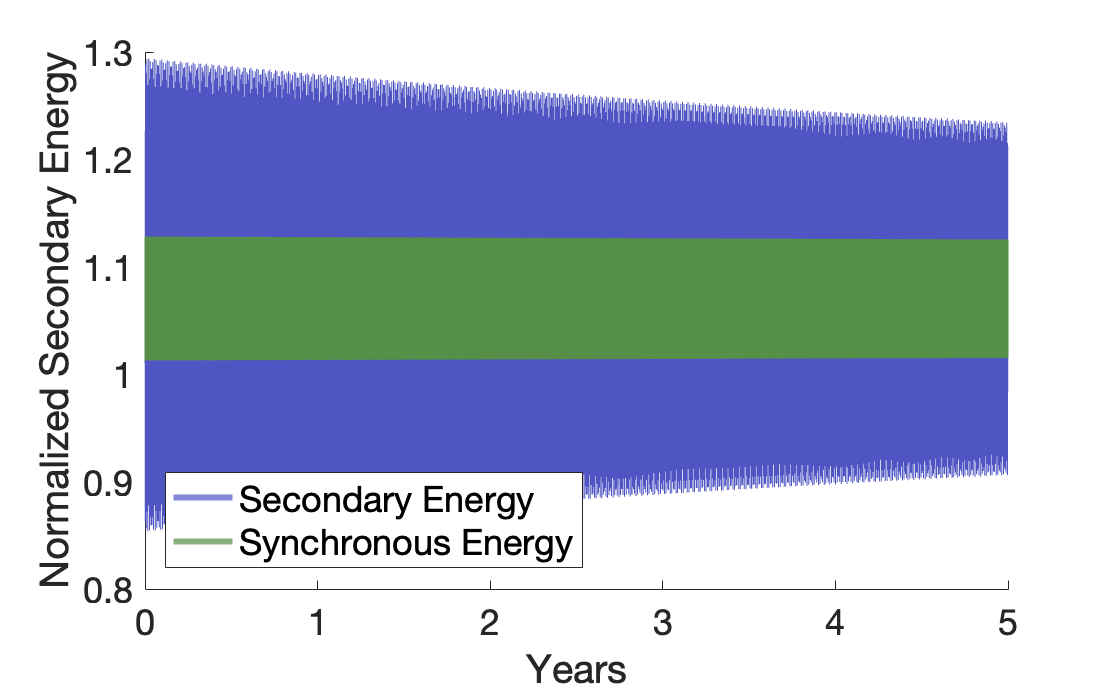} 
  \caption{The secondary rotational energy for the stable system ($a/b=1.2$, $b/c=1.1$) in blue, along with the synchronous energy in green. The energy is normalized by the pre-impact equilibrium energy. The instantaneous energy approaches the synchronous value as energy dissipates.}
  \label{sec_energy_stable}
\end{figure}

As seen in Fig. \ref{energy_stable}, the secondary energy does not initially evolve secularly but rather damps oscillations to approach a synchronous state. We can calculate the synchronous energy the secondary would have if it were in an equilibrium using the synchronous spin rate of the system calculated by \cite{scheeres2009stability}:
\begin{equation}
    \dot{\theta}^{2}=\frac{\mathcal{G}(M_A+M_B)}{r^3}\bigg[1+\frac{3}{2r^2}\bigg(\bar{I}_{B,z}+\bar{I}_{B,y}-2\bar{I}_{B,x}\bigg)\bigg]
\end{equation}
where the bodies are separated by a distance of $r$, and the bar indicates the mass-normalized secondary inertia values. Using this equilibrium spin rate, the secondary's minimum energy state at any time is simply calculated as:
\begin{equation}
    E_{min}=\frac{1}{2}I_{B,z}\dot{\theta}^{2}.
\end{equation}
Fig. \ref{sec_energy_stable} shows the secondary's rotational energy along with its instantaneous synchronous energy over the first 5 years for the stable system ($a/b=1.2$, $b/c=1.1$). 

\begin{figure}[ht!]
  \centering
  \includegraphics[width = 3in]{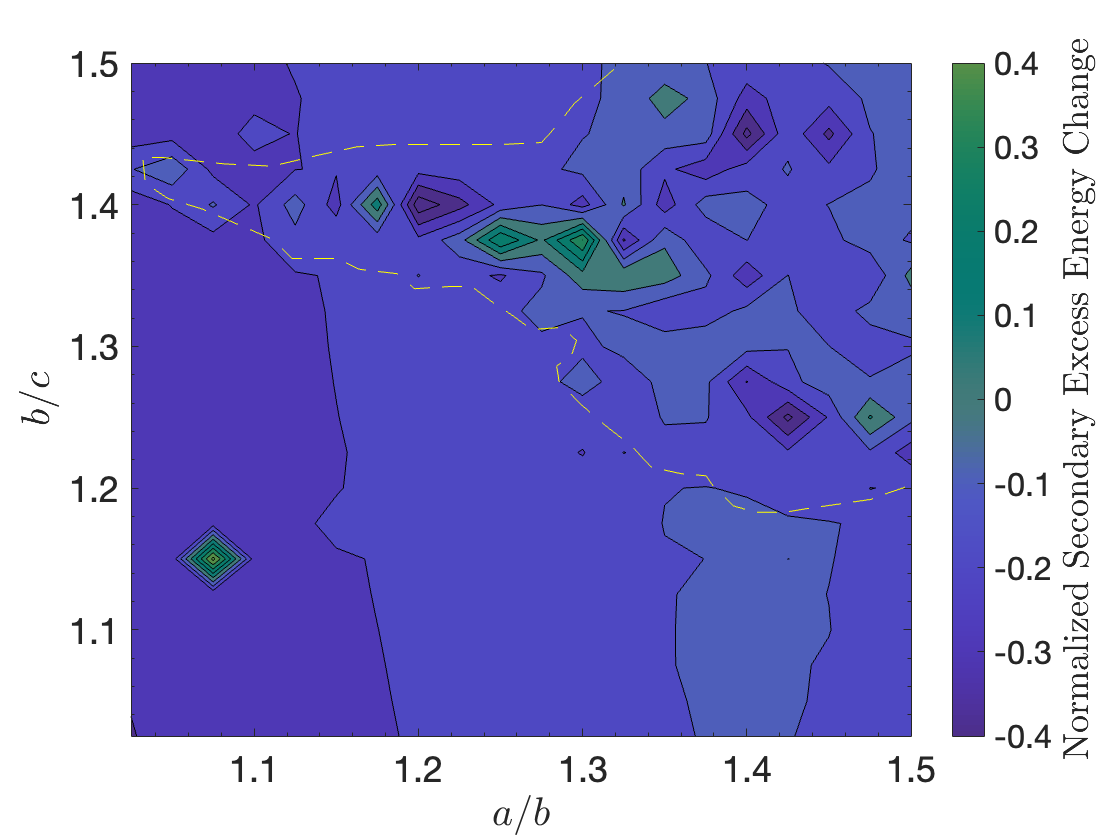} 
  \caption{The normalized change in secondary excess energy after 5 years, calculated as the difference between the average secondary excess energy during the first and last year of the simulation, normalized by the average excess energy during the first year of the simulation. The yellow dashed line shows the approximate unstable region.}
  \label{EB_delta}
\end{figure}

We see the secondary energy approaching the synchronous state as the system dissipates energy. To quantify this we define the secondary excess energy, which is simply
\begin{equation}
    E_{\text{exc}}=|E_{\text{B}}-E_{\text{min}}|.
\end{equation}
By calculating changes in the excess energy, we can determine how close the system has moved to the synchronous configuration. We investigate the secondary's energy change for all shapes by plotting the relative change in excess energy over 5 years in Fig. \ref{EB_delta}. The relative change is calculated as
\begin{equation}
    \Delta E = \frac{\bar{E}_5-\bar{E}_1}{\bar{E}_1}.
\end{equation}
where $\bar{E}_5$ is the mean excess energy over the fifth year of the simulation, and likewise for $\bar{E}_1$. In other words, the relative change is the difference in the mean excess energy over the first and fifth (final) year of the simulation, normalized by the average excess energy during the first year. In Fig. \ref{EB_delta}, we see essentially random changes in the secondary excess energy for shapes within the unstable region (refer to Fig. \ref{beta3_impact} for an illustration of the unstable region, but generally this is in the region of high values of $a/b$ and $b/c$). Due to the chaotic dynamics of these shapes, the secondary could actually be moving away from the synchronous state. Outside the unstable region, generally we see shapes with a small value of $a/b$ have the largest relative change in excess energy, while secondaries with $a/b\approx1.4$ have the smallest relative change. Interestingly, there is a large resonance in the system when $a/b\approx1.4$ \citep{agrusa2021excited}, which leads to a more energetic response in the secondary (higher initial libration amplitude; see Fig. \ref{Lib_max} below), so the relative change in secondary energy is smaller.

\subsection{Free Energy}
\begin{figure}[ht!]
   \centering
   \includegraphics[width = 3in]{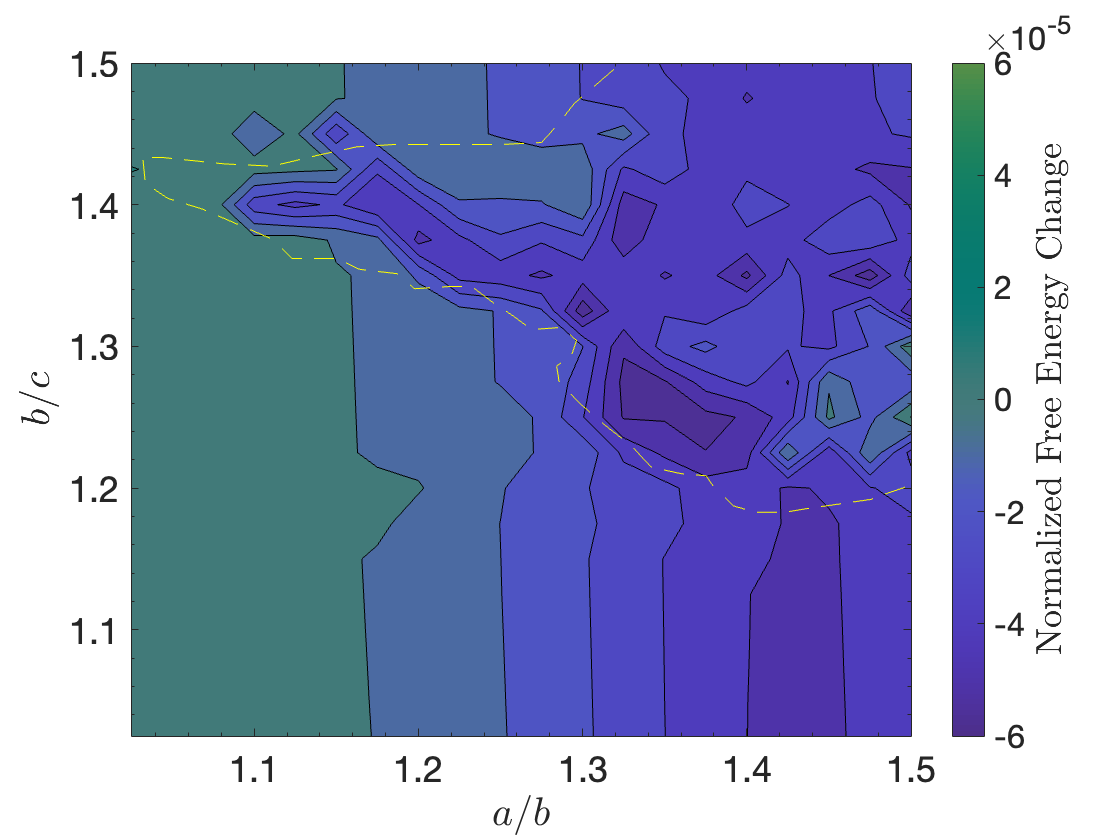} 
   \caption{The percent change in free energy after 5 years, calculated as the difference between the average free energy during the first and last year of the simulation, normalized by the average energy during the first year of the simulation. The yellow dashed line shows the approximate unstable region.}
   \label{EF_delta}
\end{figure}

We next turn to the free energy, which is the sum of the secondary and orbit energies. By excluding the primary's energy, this allows us to better understand the system's evolution. We keep the same definition for the relative change in energy, but now applied to the free energy. The relative change in free energy is plotted in Fig. \ref{EF_delta}, where we see in general very small changes in the first 5 years. However, as discussed above, in some systems the free energy is initially decreasing. In Fig. \ref{EF_delta} we see this is the case for most systems, as only shapes with a small value of $a/b$ immediately start increasing free energy. We point out that as the orbit expands (semimajor axis increases), we would expect the free energy to also increase as the orbit energy increases, yet this is not the case during the first years after the perturbation. This once again highlights the departure from classical tidal theory during the initial period of libration damping as dissipation pushes the system back toward an equilibrium. Again we see the effect of the resonance at $a/b=1.4$, as shapes near this area (and outside the unstable region) have the largest relative decrease in free energy over the first 5 years. Within the unstable region generally the free energy is decreasing, but this is not true for every shape, illustrating the chaotic nature of this region.

\subsection{Libration}
\begin{figure}[ht!]
   \centering
   \includegraphics[width = 3in]{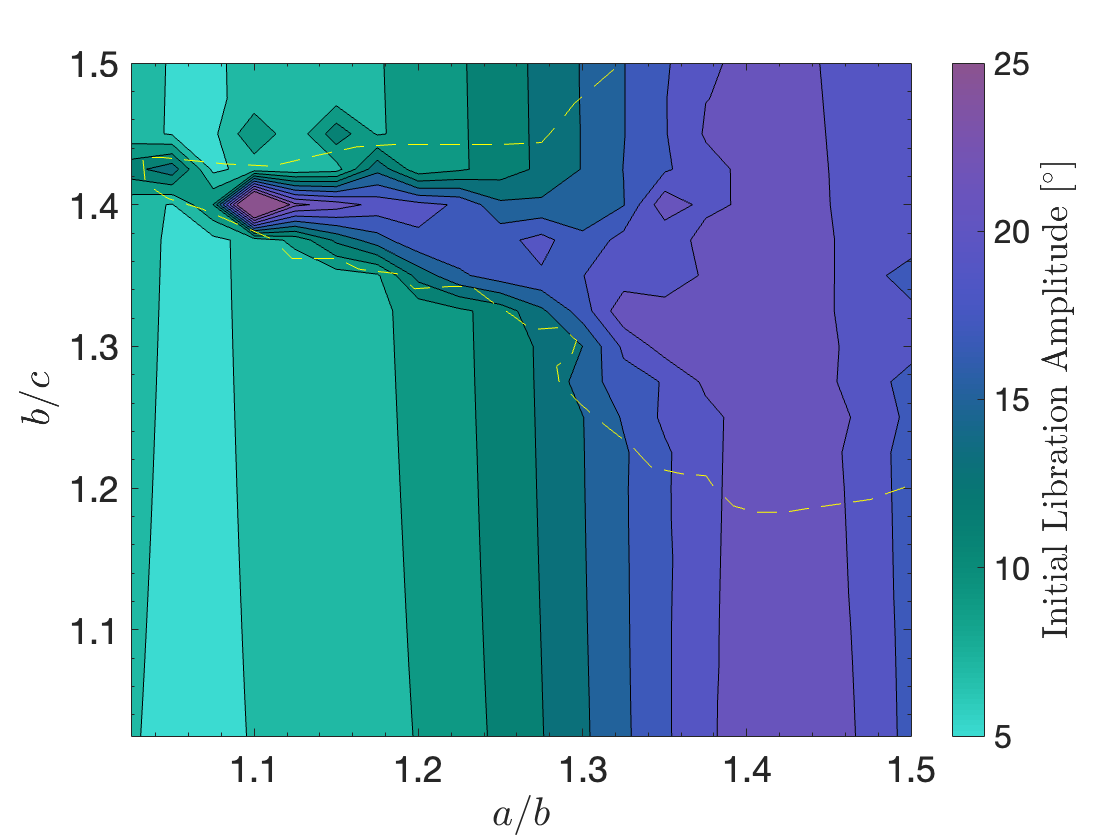} 
   \caption{The initial libration amplitude at the start of the simulation. The yellow dashed line shows the approximate unstable region, which generally has the largest libration amplitudes.}
   \label{Lib_max}
\end{figure}

Of particular interest is the libration amplitude of the system, as this will be a physically observable quantity with Hera. Here, we are defining libration as the angle between the secondary's long axis and the radial vector from the secondary to the primary. Thus, a system in equilibrium will have zero libration as the secondary's long axis is exactly aligned with the primary. In Fig. \ref{Lib_max}, we plot the initial libration amplitude after the impact. Generally, we see the libration amplitude increases with the secondary's elongation, but begins to drop off after the resonance near $a/b=1.4$, which is where the libration amplitude is at a maximum. Outside the unstable region, libration amplitude is essentially independent of $b/c$, but within the unstable region we generally see larger libration amplitudes as a result of the out-of-plane secondary rotation.

There is a correlation between the initial libration amplitude and the initial free energy trend, which can be seen when comparing Figs. \ref{EF_delta} and \ref{Lib_max}. Generally, a larger libration amplitude corresponds to a faster decrease in free energy, while systems with a small libration amplitude immediately start increasing free energy. This indicates that an initial decrease in free energy could be caused by a large perturbation away from equilibrium, as dissipation forces the system back toward an equilibrium configuration.

\begin{figure}[ht!]
   \centering
   \includegraphics[width = 3in]{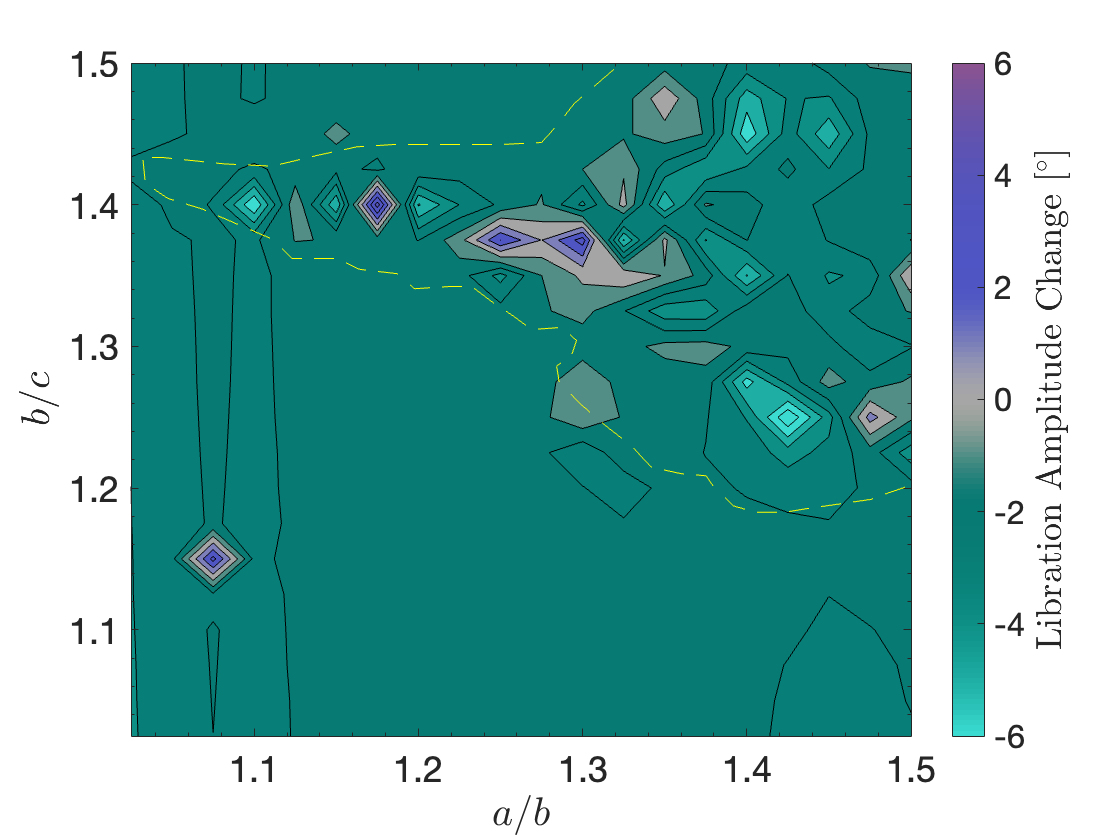} 
   \caption{The change in libration amplitude after 5 years, calculated as the difference between the maximum libration during the first and last year of the simulation. The yellow dashed line shows the approximate unstable region, in which the libration amplitude can change randomly.}
   \label{Lib_delta}
\end{figure}

Fig. \ref{Lib_delta} plots the change in libration amplitude after 5 years. This is calculated by finding the difference between the maximum libration angles between the first and last year of the simulation. Within the unstable region, the system can increase or decrease the libration amplitude over this short time span depending on the shape. Thus, if the DART impact causes Dimorphos to become attitude unstable, Hera won't observe any systematic trend in libration amplitude and instead see chaotic variations due to the non-principal axis rotation. However, outside the unstable region there is a systematic decrease in libration amplitude, although only a few degrees for our selected values of $Q/k$. However, the largest decrease in libration amplitude outside the chaotic region is around $a/b=1.1$, which is another resonance \citep{agrusa2021excited}. These systems could see a decrease in libration amplitude of around 3-4$^\circ$, which is noteworthy considering these same systems have a small initial libration amplitude of around 8-10$^\circ$. Thus, it's possible Dimorphos could dissipate a significant fraction of the DART-induced libration amplitude by the time of Hera's arrival, of course depending on the true tidal parameters of the system. A discussion of the effect of $Q/k$ on libration damping is included in Section \ref{sec:QK}.

\subsection{Eccentricity}
\begin{figure}[ht!]
   \centering
   \includegraphics[width = 3in]{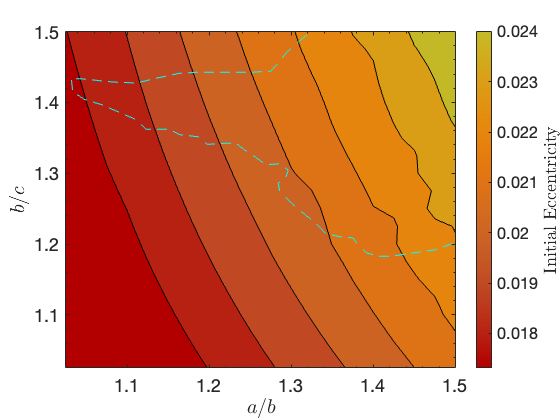} 
   \caption{The initial eccentricity, calculated as the maximum eccentricity during the first year of the simulation. The cyan dashed line shows the approximate unstable region, which has no bearing on the initial eccentricity.}
   \label{ecc_max}
\end{figure}

\begin{figure}[ht!]
   \centering
   \includegraphics[width = 3in]{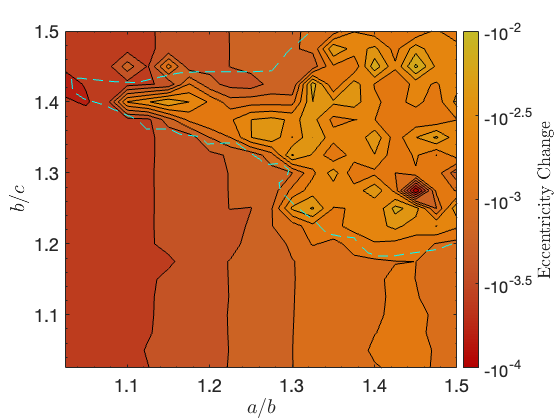} 
   \caption{The change in eccentricity after 5 years, calculated as the difference between the maximum eccentricity during the first and last year of the simulation. The results are plotted on a logarithmic scale for clarity. The cyan dashed line shows the approximate unstable region, which sees the largest change in eccentricity.}
   \label{ecc_delta}
\end{figure}

Next we examine the Keplerian eccentricity of the system. Fig. \ref{ecc_max} plots the maximum eccentricity over the first year after the perturbation, where we see shapes with large values of $a/b$ and $b/c$ have the largest initial eccentricity. There appears to be no effect from the unstable region on the initial eccentricity. However, this is not the case for the change in eccentricity over 5 years, shown in Fig. \ref{ecc_delta}. Here we see the largest decreases in eccentricity are found for shapes in the unstable region, consistent with the findings of \cite{quillen2020excitation}. Outside the unstable region, as $a/b$ increases so does the change in eccentricity. Thus, it appears the shapes with the largest initial libration amplitudes also see the largest decrease in eccentricity during the first years after the perturbation. These are also the same shapes with the largest initial decrease in free energy, but the smallest relative change in secondary excess energy. This highlights the importance of spin-orbit coupling, as systems pushed furthest away from equilibrium see the fastest changes to their orbit in order to bring the secondary's spin back to an equilibrium.

\section{BYORP} \label{sec:byorp} %###################################################
%###################################################%###################################################
While this analysis has focused on the internal system dynamics of a binary asteroid, an important external influence on the dynamics is BYORP \citep{cuk2005effects}. For a singly synchronous binary system, BYORP can act to either expand or contract the orbit, with NPA rotation decreasing the magnitude of the BYORP torque \citep{quillen2022non}. Thus, we expect the BYORP effect to have a smaller impact on the dynamics of the unstable systems where NPA rotation is frequent.

\cite{jacobson2011long} predict an equilibrium in which tidal dissipation balances BYORP drift, and the orbit does not evolve over time. Didymos has a very small mean anomaly drift rate of $\Delta M=0.15\pm0.14^\circ yr^{-2}$ with $3\sigma$ uncertainty \citep{scheirich2022preimpact}, meaning that if it is not in a tide-BYORP equilibrium, it is likely near one. For our purposes we will assume an equilibrium between tides and BYORP in order to calculate a BYORP coefficient as a check on our analysis. Given the small observed mean anomaly drift, we can make this assumption without introducing too much error. In such an equilibrium, we can calculate:
\begin{equation}
    \frac{BQ_A}{k_A}=\frac{2\pi\omega_d^2\rho R_A^2q^{4/3}}{H_\odot a^7}
    \label{byorp_equation}
\end{equation}
where $\omega_d=\sqrt{4\pi\mathcal{G}\rho/3}$ is the spin disruption limit, $q=M_B/M_A$ is the mass ratio (under a uniform system bulk density assumption, this is equivalent to a volume ratio), $R_A$ and $a$ are the primary's mean radius and the binary mutual semimajor axis, respectively, and $H_\odot$ is a solar parameter defined as
\begin{equation}
    H_\odot = \frac{F_s}{a_\odot^2\sqrt{1-e_\odot^2}}
\end{equation}
where $F_s\approx1\times10^{17}$ is a constant and $a_\odot$, $e_\odot$ are the heliocentric semimajor axis and eccentricity of Didymos, respectively \citep{mcmahon2010detailed}.

Likely, magnitudes of $B$ exist in the interval from 0 to $10^{-2}$, but most commonly values of $B$ are reported between $10^{-3}$ and $10^{-2}$ \citep{scheirich2015binary, jacobson2011long, steinberg2011binary}. With our parameters for Didymos and our value for $Q_A/k_A$, we calculate $B\approx4.5\times10^{-3}$, which is within the expected interval.

\begin{figure}[ht!]
   \centering
   \includegraphics[width = 3in]{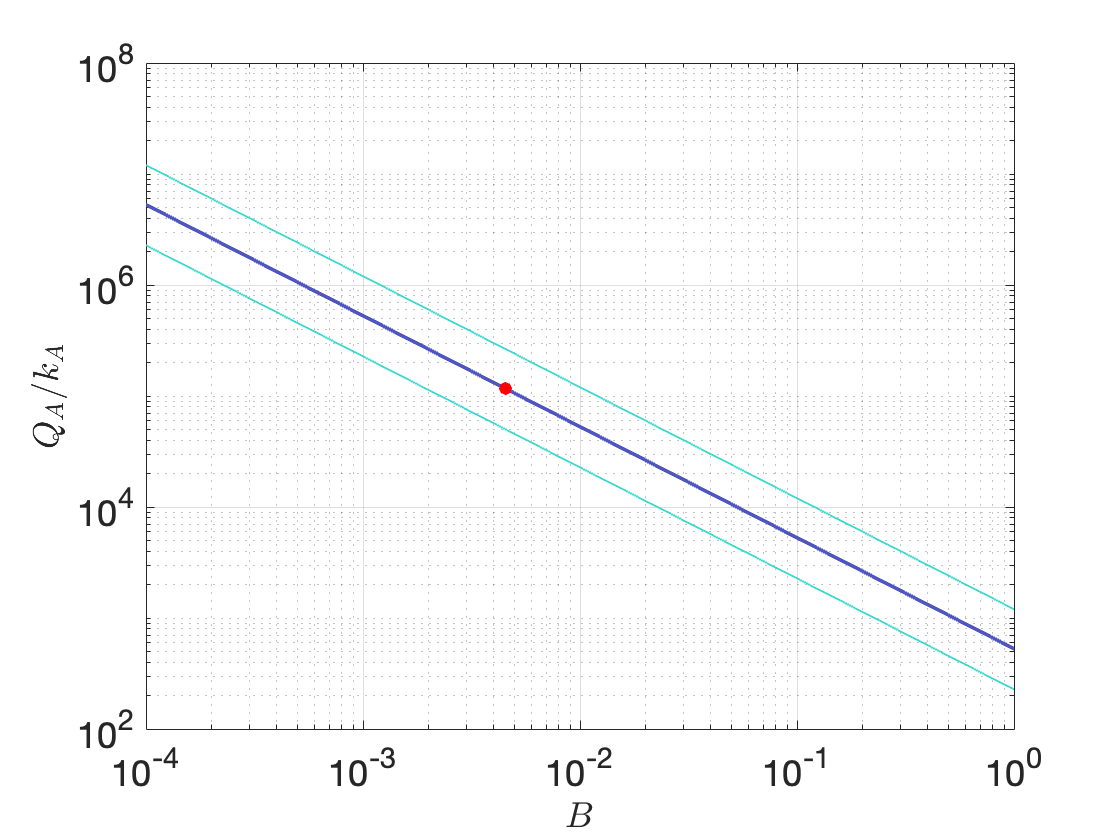} 
   \caption{$Q_A/k_A$ as a function of the BYORP coefficient $B$ for our nominal Didymos system, plotted in dark blue. The uncertainty envelope around this curve is plotted in cyan considering the error bars in the bodies' estimates of sizes and densities. Our estimate is shown as a red dot on the plot.}
   \label{BQk_plot}
\end{figure}

Using Eq. \ref{byorp_equation}, we can calculate the relationship between $B$ and $Q_A/k_A$ for our nominal Didymos system, shown in Fig. \ref{BQk_plot}, for a BYORP-tide equilibrium. There is additional uncertainty around the size of both Didymos and Dimorphos, as well as the system bulk density which we have calculated independently using the system dynamics. Using the error bars on the size and density of these bodies, we also plot an uncertainty envelope around the $BQ_A/k_A$ curve in Fig. \ref{BQk_plot}. 

On Fig. \ref{BQk_plot}, we plot our nominal estimate for $B$ that matches the $Q_A/k_A$ value we adopted as a red dot. While this corresponds to a reasonable value of $B$, there is still a range of possibilities that should be considered. For example, for values of $B$ within $10^{-3}-10^{-2}$, $Q_A/k_A$ could vary by nearly an order of magnitude in either direction. While the consistency seen here lends confidence to our selection of $Q_A/k_A$, we will investigate the role different values of $Q/k$ have on energy dissipation, for both the primary and secondary.

In Didymos, BYORP acts to contract the orbit (decrease the semimajor axis) \citep{scheirich2022preimpact}. However, from our analysis, the time for eccentricity and libration to damp to nearly zero is relatively fast and the semimajor axis changes on the order of only 1 meter during this time (see Figs. \ref{elements_stable} and \ref{elements_unstable}). Thus, during the timescale we are interested in for this analysis, we expect BYORP to have very little effect on the dynamics, mainly decreasing the secular slope of semimajor axis evolution. Furthermore, \cite{quillen2022non} predict that the BYORP effect is weakened as a result of NPA rotation, so unstable systems would be affected even less.

The BYORP coefficient is primarily a function of Dimorphos' shape, thus the impact and subsequent reshaping of Dimorphos will change the coefficient at some level. Previous work has shown that significant reshaping of Dimorphos is possible \citep{hirabayashi2022double, nakano2022nasa, raducan2022global}, so it is difficult to predict what level of change there will be in the BYORP coefficient. For a small change where BYORP remains contractive, Eq. \ref{byorp_equation} indicates that the shift in equilibrium will be small. However if $B$ changes significantly, there could be a significant change to the equilibrium semimajor axis (if BYORP is contractive) or BYORP could act expansively along with tides to grow the orbit over time. In either case, however, even a significantly higher BYORP rate in either direction will have minimal effect on the orbit over the coming decades.

\section{Effect of Tidal Parameters} \label{sec:QK} %###################################################
%###################################################%###################################################
As previously mentioned, there is considerable uncertainty surrounding the tidal parameters $Q/k$ for both the primary and secondary. In Section \ref{sec:byorp} we saw $Q_A/k_A$ can vary by an order of magnitude in either direction and still maintain consistency with a reasonable BYORP coefficient and the uncertainty around the shape and density of Didymos. Thus, we define three values for $Q_A/k_A$ to test: $10^4$, $10^5$, and $10^6$. We use these same values to test $Q_B/k_B$ as well. First, we vary $Q_A/k_A$ over these three values while holding $Q_B/k_B=10^5$ constant. Then we perform the opposite test by varying $Q_B/k_B$ over the same values while holding $Q_A/k_A$ constant. In this way we can determine which behaviors in the system rely mainly on the primary or secondary. Given how quickly dissipation occurs in our analysis in Section \ref{sec:Dissipation}, we do not think it likely that $Q/k$ is below $10^4$, and even if it is this would only speed up the process already observed. Furthermore, it is possible that $Q/k$ is larger than $10^6$, but if this is the case, again the system would only evolve more slowly than the $10^6$ cases we test below. Thus, this range of $Q/k$ values for both bodies gives us an idea on how the tidal parameter affects the system evolution. We perform this analysis for both the stable and unstable systems.

\subsection{Stable System}
\begin{figure*}[ht!]
   \centering
   \includegraphics[width = 6.5in]{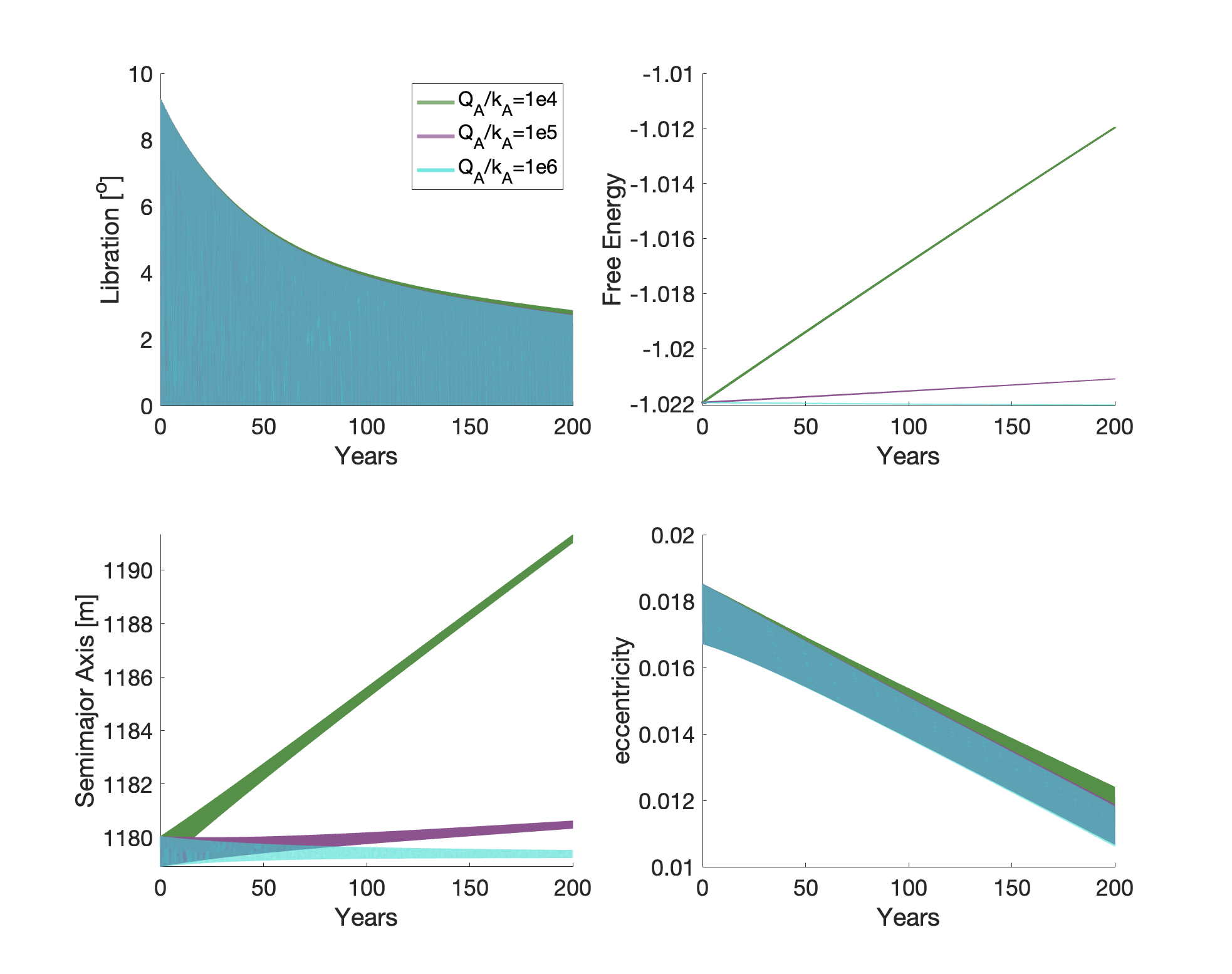} 
   \caption{For the stable system $a/b=1.2$, $b/c=1.1$, we vary the value of $Q_A/k_A$ while keeping $Q_B/k_B=10^5$ constant. For 200 years, we plot the libration angle (top left), normalized free energy (top right), semimajor axis (bottom left), and eccentricity (bottom right). While the rate of change of the semimajor axis and free energy strongly depend on $Q_A/k_A$, the libration amplitude and eccentricity are largely independent of primary tidal parameters.}
   \label{QKprim_stable}
\end{figure*}

For the stable system ($a/b=1.2$, $b/c=1.1$), we first hold $Q_B/k_B=10^5$ constant and vary $Q_A/k_A$ between $10^4$ and $10^6$. In Fig. \ref{QKprim_stable}, we plot the libration amplitude, free energy, semimajor axis, and eccentricity over 200 years. The free energy has been normalized by its pre-impact equilibrium value. Note this is not a long enough time span for the system to fully equilibrate, but it is long enough to see the secular behavior of the system. 

From Fig. \ref{QKprim_stable}, we see the dissipation of the libration amplitude has very little dependence on $Q_A/k_A$, indicating it depends almost solely on $Q_B/k_B$. Conversely, the semimajor axis has a strong dependence on $Q_A/k_A$, as evidenced in the plot of free energy and semimajor axis. These dependencies are expected from classical tidal theory. Unsurprisingly, smaller $Q_A/k_A$ values (more dissipative systems) expand the orbit more rapidly than large values of $Q_A/k_A$. For small $Q_A/k_A$ values, the secular trend is faster than the dissipation in semimajor axis oscillations, whereas large $Q_A/k_A$ values see the opposite, where oscillations in semimajor axis are damped faster than the secular trend becomes dominant. In the plot of free energy, we see the least dissipative system slowly losing free energy while the other systems are monotonically increasing, which is consistent with the behavior seen in the semimajor axis. Overall this indicates that when $Q_A/k_A$ is larger than $Q_B/k_B$, the system contracts its orbit first to damp semimajor axis oscillations and libration amplitude, whereas when $Q_A/k_A$ is smaller than $Q_B/k_B$, the orbit expands faster than the secondary re-equilibrates. Lastly, the eccentricity damping also appears to have only a small dependence on $Q_A/k_A$, as there are only small differences in the trend between the values tested here.

\begin{figure*}[ht!]
   \centering
   \includegraphics[width = 6.5in]{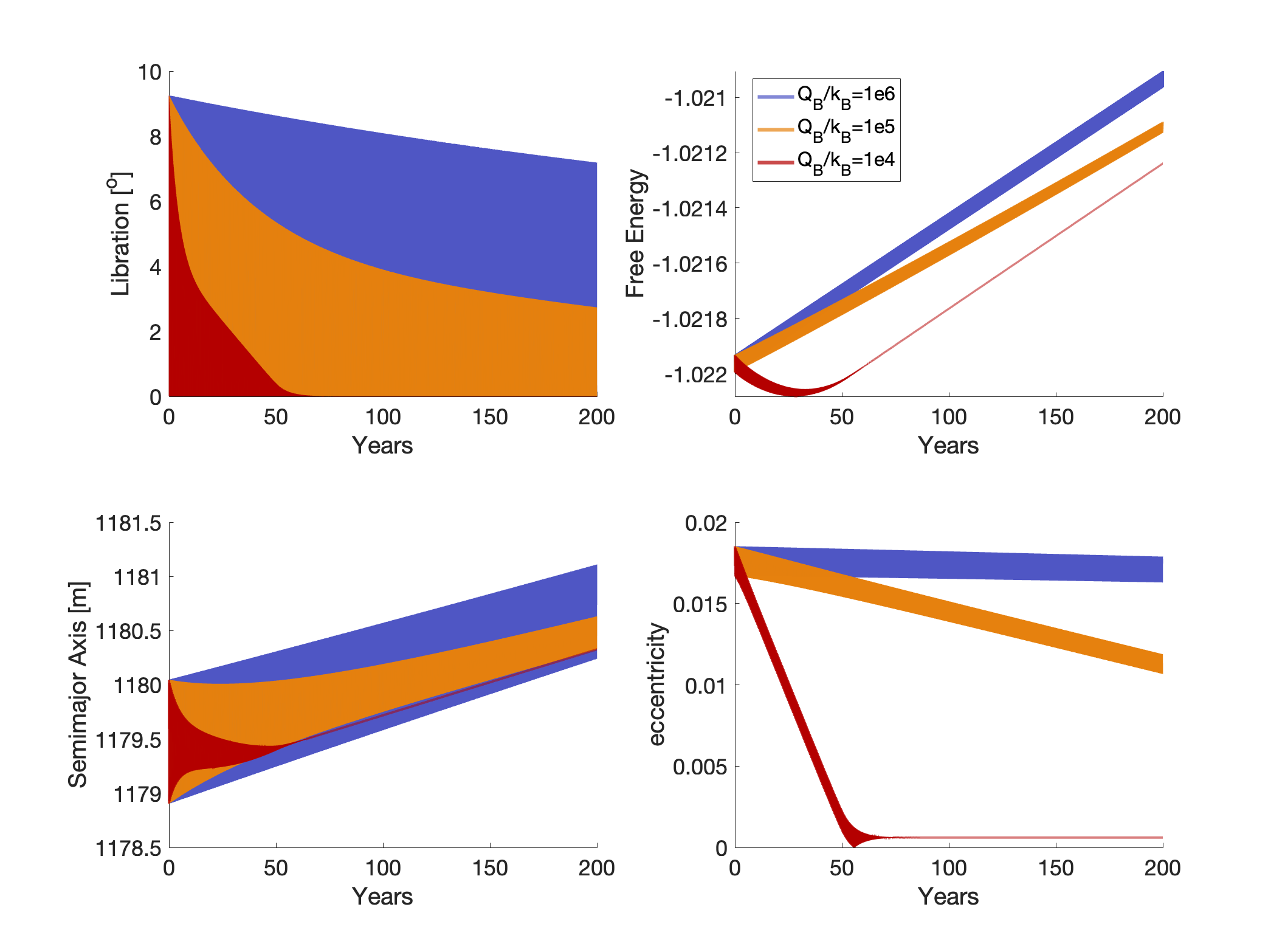} 
   \caption{For the stable system $a/b=1.2$, $b/c=1.1$, we vary the value of $Q_B/k_B$ while keeping $Q_A/k_A=10^5$ constant. For 200 years, we plot the libration angle (top left), normalized free energy (top right), semimajor axis (bottom left), and eccentricity (bottom right). The dissipation rates of the libration amplitude and eccentricity strongly depend on $Q_B/k_B$. For very dissipative secondaries, the free energy initially decreases, which is not the case for secondaries with larger $Q_B/k_B$ values. The secular rate of semimajor axis expansion seems independent of $Q_B/k_B$, but the damping rate of oscillations in semimajor axis does depend on $Q_B/k_B$.}
   \label{QKsec_stable}
\end{figure*}

Next, we perform the complement of this analysis by holding $Q_A/k_A=10^5$ constant and varying $Q_B/k_B$ between $10^4$ and $10^6$. In Fig. \ref{QKsec_stable}, we plot the libration amplitude, the free energy, semimajor axis, and eccentricity over 200 years.

In Fig. \ref{QKsec_stable}, we see the dissipation of the libration amplitude has a strong dependence on $Q_B/k_B$, with smaller values (more dissipative) damping libration faster. Looking at the free energy, the secular trend appears to have only a small dependence on $Q_B/k_B$, but for small values of $Q_B/k_B$ we see the initial decrease in free energy. While the case $Q_B/k_B=Q_A/k_A=10^5$ appears to have a slope different from the other cases, its slope is actually changing slowly and approaching the same rate as the others. The free energy corresponds to the semimajor axis, where the secular trend again sees only a small dependence on $Q_B/k_B$, but the rate of damping oscillations does have a strong dependence on $Q_B/k_B$. The most dissipative system ($Q_B/k_B=10^4$) damps the oscillations fastest and appears to have an initial trend of decreasing semimajor axis (consistent with the decrease in free energy). This indicates that systems with a very dissipative secondary initially contract the orbit to reestablish equilibrium. Lastly, we see a strong dependence of eccentricity damping on $Q_B/k_B$, with more dissipative systems unsurprisingly damping eccentricity the fastest. Again, these dependencies are expected from classical tidal theory.

\subsection{Unstable System}
\begin{figure*}[ht!]
   \centering
   \includegraphics[width = 6.5in]{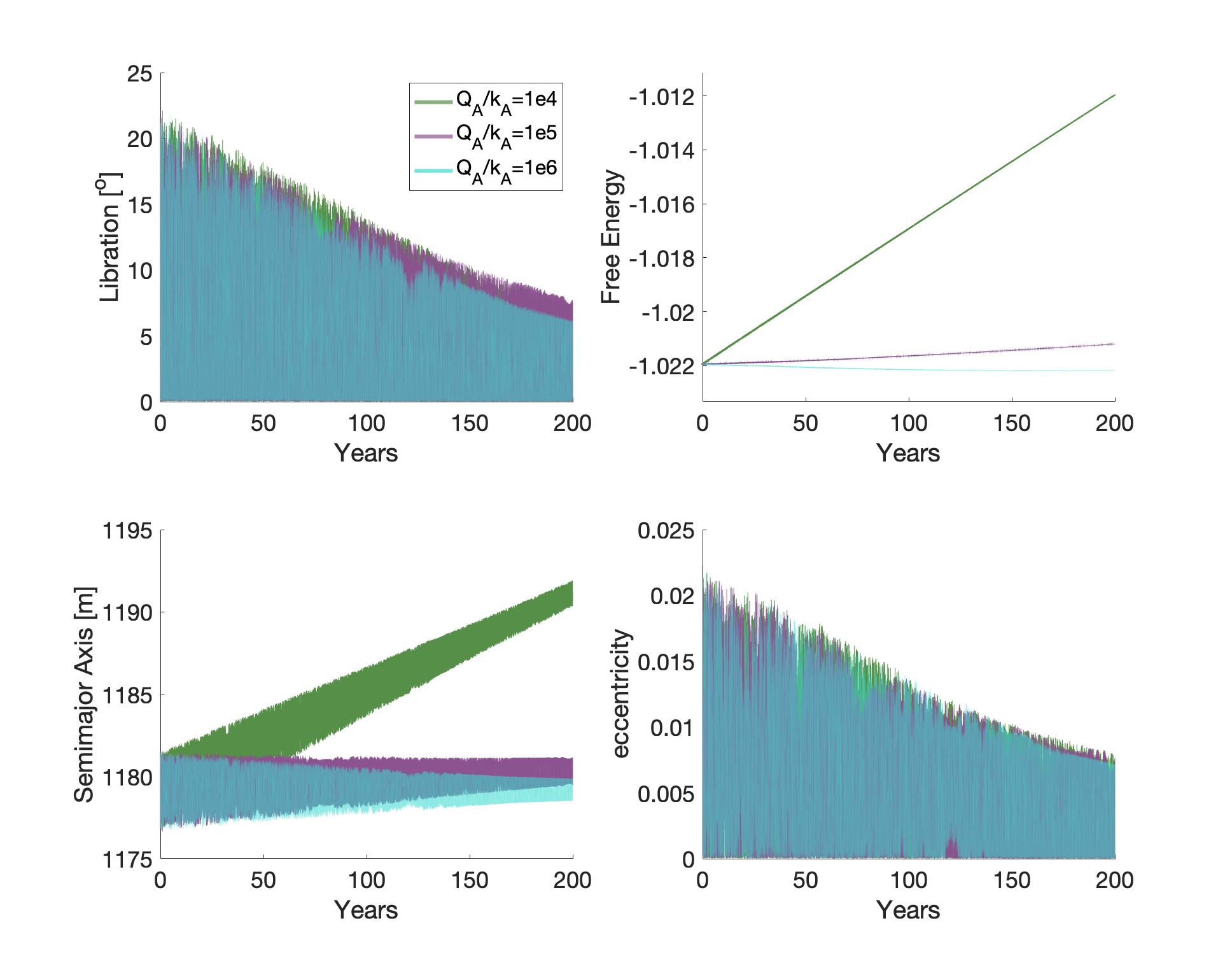} 
   \caption{For the unstable system $a/b=1.4$, $b/c=1.3$, we vary the value of $Q_A/k_A$ while keeping $Q_B/k_B=1\times10^5$ constant. For 200 years, we plot the libration angle (top left), normalized free energy (top right), semimajor axis (bottom left), and eccentricity (bottom right). While the rate of change of the semimajor axis and free energy strongly depend on $Q_A/k_A$, the libration amplitude and eccentricity are largely independent of primary tidal parameters.}
   \label{QKprim_unstable}
\end{figure*}

We next repeat the same analysis by varying $Q_A/k_A$ and $Q_B/k_B$ for the unstable system ($a/b=1.4$, $b/c=1.3$). First, we hold $Q_B/k_B=10^5$ constant while varying $Q_A/k_A$ between $10^4$ and $10^6$. We plot the libration amplitude, free energy, semimajor axis, and eccentricity for this analysis in Fig. \ref{QKprim_unstable}.

Overall, we see very similar behavior between the unstable and stable system. Consistent with classical tidal theory, it appears the libration amplitude and eccentricity damping are largely unaffected by the value of $Q_A/k_A$ during this time. Again, the secular trend of free energy and semimajor axis strongly depend on $Q_A/k_A$ as expected. When $Q_A/k_A>Q_B/k_B$ (i.e. $Q_A/k_A=10^6$, $Q_B/k_B=10^5$), the damping rate of semimajor axis oscillations is faster than the secular trend, and as a result we see an overall decrease in the free energy. At a longer timescale we expect both the semimajor axis and free energy to begin increasing as the orbit expands.

\begin{figure*}[ht!]
   \centering
   \includegraphics[width = 6.5in]{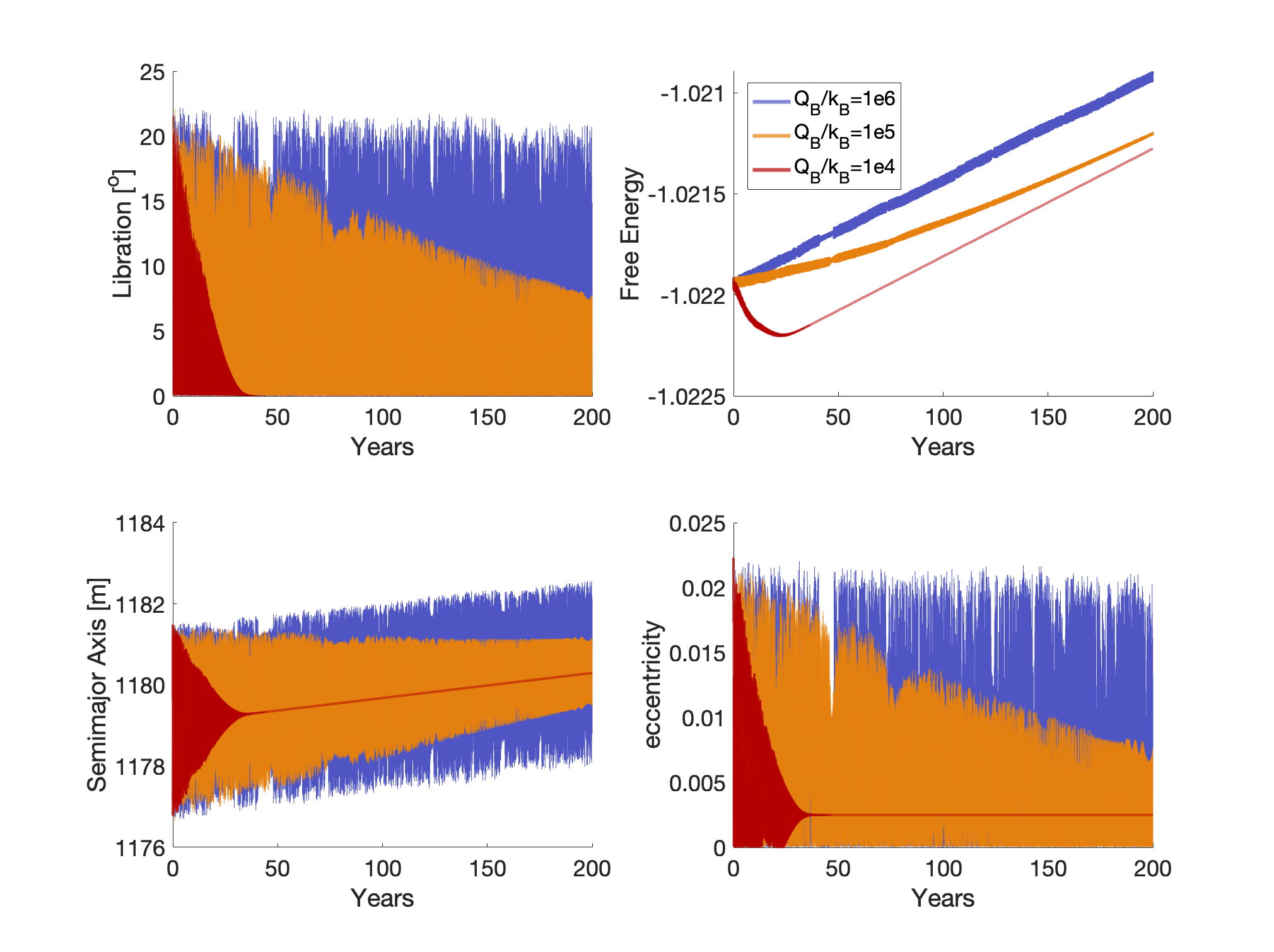} 
   \caption{For the unstable system $a/b=1.4$, $b/c=1.3$, we vary the value of $Q_B/k_B$ while keeping $Q_A/k_A=1\times10^5$ constant. For 200 years, we plot the libration angle (top left), normalized free energy (top right), semimajor axis (bottom left), and eccentricity (bottom right). The dissipation rates of the libration amplitude and eccentricity strongly depend on $Q_B/k_B$. For very dissipative secondaries, the free energy initially decreases, which is not the case for secondaries with larger $Q_B/k_B$ values. The secular rate of semimajor axis expansion seems independent of $Q_B/k_B$, but the damping rate of oscillations in semimajor axis does depend on $Q_B/k_B$.}
   \label{QKsec_unstable}
\end{figure*}

We next hold $Q_A/k_A=10^5$ constant and vary $Q_B/k_B$ between $10^4$ and $10^6$. For this test, the libration amplitude, free energy, semimajor axis, and eccentricity are plotted in Fig. \ref{QKsec_unstable}.

Again, there are strong similarities between the unstable and stable case. The damping rate of libration amplitude and eccentricity strongly depend on $Q_B/k_B$, while the secular trends in semimajor axis and free energy seem ignorant of $Q_B/k_B$, as expected from classical tidal theory. However, the damping rate of semimajor axis oscillations does depend heavily on $Q_B/k_B$, with more dissipative secondaries (small $Q_B/k_B$) damp these oscillations faster than the secular trend develops, and as a result the free energy of the system initially decreases.

Based on these analyses of $Q_A/k_A$ and $Q_B/k_B$, we can conclude that the libration amplitude, eccentricity, and oscillations in semimajor axis mostly depend on $Q_B/k_B$, while the secular trend of the orbit, i.e. semimajor axis expansion, is driven by $Q_A/k_A$. When systems have a more dissipative secondary, they re-enter equilibrium before any noticeable secular change in the orbit develops, whereas when systems have a more dissipative primary the re-equilibrization of the secondary takes longer than the secular evolution of the orbit.

\section{Discussion} \label{sec:discussion} %###################################################
%###################################################%###################################################
In this work we attempt to outline the possibilities of energy dissipation after a DART-like perturbation both in the long- and near-term, investigating dissipation both in stable and tumbling systems. We implement tidal torque and non-principal axis dissipation in a simple sphere-ellipsoid approximation of binary asteroids. We find that both stable and tumbling systems dissipate energy on comparable timescales to return to a synchronous configuration. Previous studies have claimed that tumbling greatly reduces the rate of energy dissipation \citep{wisdom1984chaotic, naidu2015near, quillen2022non}, but this is only for the non-synchronous case. Our results agree with \cite{quillen2020excitation} in that tumbling within the synchronous state can increase energy dissipation, and the libration amplitude damps to zero before predicted for planar rotation. A unique result we find is that non-principal axis rotation can damp as quickly as planar libration for strongly coupled systems with efficient dissipation. We find in these systems the libration amplitude, both stable and unstable, is tied closely to the orbit eccentricity and oscillations in the semimajor axis, as all of these dissipate on the same timescale. For especially dissipative systems, the system returns to equilibrium before any substantial secular trend is apparent.

For near-term dynamics relevant to the Hera mission, we find systems experiencing stable in-plane libration systematically dissipate energy to return to an equilibrium, synchronous configuration. However, systems with out-of-plane tumbling do not have a systematic trend during the short time between DART's impact and Hera's rendezvous thanks to chaotic dynamics. Thus, the shape of Dimorphos is paramount in predicting energy dissipation, as not only does the shape place the system in the stable/unstable region, it also dictates the magnitude of post-impact libration. Generally, more elongated shapes and shapes in the unstable region have the largest libration amplitude and the largest relative change in free energy following the impact. However, the secondary excess energy can vary randomly within the unstable region, and in the stable region less elongated shapes (i.e. shapes with a smaller initial libration amplitude) have the largest relative decrease in secondary excess energy. The rate of libration damping also depends on the shape, as unstable shapes again can see random changes in the libration amplitude. In the stable region, resonances play a large role in how the libration amplitude dissipates, but in general this is not expected to exceed a few degrees in the years between the DART impact and Hera's arrival, even in the most dissipative cases. However, there are some cases, with a very dissipative secondary, in which a small initial libration amplitude can decrease by 20-30$\%$. The largest decrease in eccentricity is found for shapes with the largest initial libration amplitude, but overall this change is still small, on the order of 0.01 over 5 years for very dissipative systems. Thus, there is potential for non-negligible changes to occur in the system before Hera's arrival, contingent on the secondary's stiffness.

While BYORP will factor in to the evolution of the system, we believe it will have a negligible effect on the actual libration damping. In Didymos, BYORP contracts the orbit, or shrinks the semimajor axis \citep{scheirich2022preimpact}. Given the small mean anomaly drift rate, Didymos is likely near a tide-BYORP equilibrium in which the tidal expansion is nearly balanced by the BYORP effect. As a result, we can make a prediction of the BYORP coefficient, which is consistent with expected values. After the impact, the system will dissipate energy to reduce libration and eccentricity, while damping oscillations in the semimajor axis. During this time, BYORP will continue to shrink the semimajor axis, unless the impact causes significant reshaping of the secondary and changes the BYORP coefficient. However, the secular rate of semimajor axis change is already small, and generally independent of the oscillations (the secular rate is driven by the primary, whereas the oscillations are driven by the seconadry). Furthermore, NPA rotation will also decrease the BYORP effect \citep{quillen2022non}. Thus, we predict the process of libration damping will be largely independent of the secular changes caused by BYORP over the timeframe of interest here.

By varying $Q/k$ for both the primary and secondary, we find that the rate of libration and eccentricity damping are strongly dependent on the secondary's tidal parameters, but largely independent of the primary's. By extension, the oscillations in the semimajor axis also are mainly dependent on the secondary's tidal parameters. Conversely, the secular trend in the semimajor axis mainly depends on the primary's tidal parameters, but not the secondary's. Thus, we find systems with $Q_B/k_B<Q_A/k_A$ damp libration and eccentricity faster than secular changes in the orbit become apparent. This also corresponds to an initial decrease in the free energy before the semimajor axis begins expanding, unless the initial libration amplitude is small. On the other hand, when $Q_B/k_B>Q_A/k_A$, the secular trend in the orbit is immediately obvious and the damping of libration and eccentricity is relatively slow in comparison. Thus, if the secondary is very dissipative, Hera may be able to measure the damping of the libration amplitude and eccentricity. In these very dissipative systems, the coupling between eccentricity and libration means the eccentricity dissipates much faster than predicted by the analytic models of \cite{goldreich2009tidal}. Consequently, we do not recommend using an analytic model to approximate evolution of a coupled system experiencing libration, unless this system is not very dissipative or is already in an equilibrium. This also suggests close binary asteroids with measured eccentricity have either a tumbling secondary or a stiff, non-dissipative secondary, consistent with \cite{pravec2016binary}.

Future work on this topic is necessary after both the DART impact and Hera's survey. These missions will provide more information on the system, specifically the secondary shape and impact strength from DART and LICIACube and the density and constraints on the tidal parameters from Hera. With estimates on these parameters, more accurate predictions can be made on the energy dissipation from Didymos.

\section*{Acknowledgements}
The authors would like to acknowledge the DART investigation team for helpful discussion throughout this study's development and preparation. We also thank Dr. Federico Zoppetti for several useful conversations regarding the intricacies of tidal torque models. This study was supported in part by the DART mission, NASA contract No. 80MSFC20D0004 to JHU/APL. A.J.M. acknowledges support from the Planetary Defense Conference student grant. G.N. and Ö.K. acknowledge support by Belgian Federal Science Policy (BELSPO) through the ESA/PRODEX Program. R.N. acknowledges support from NASA/FINESST (NNH20ZDA001N/80NSSC22K0534).

The simulations shown in Fig. \ref{beta3_impact} were carried out on The University of Maryland Astronomy Department’s YORP cluster, administered by the Center for Theory and Computation.

\appendix
\section{Supplementary Figures}
\label{plots}

\begin{figure}[ht!]
   \centering
   \includegraphics[width = 3in]{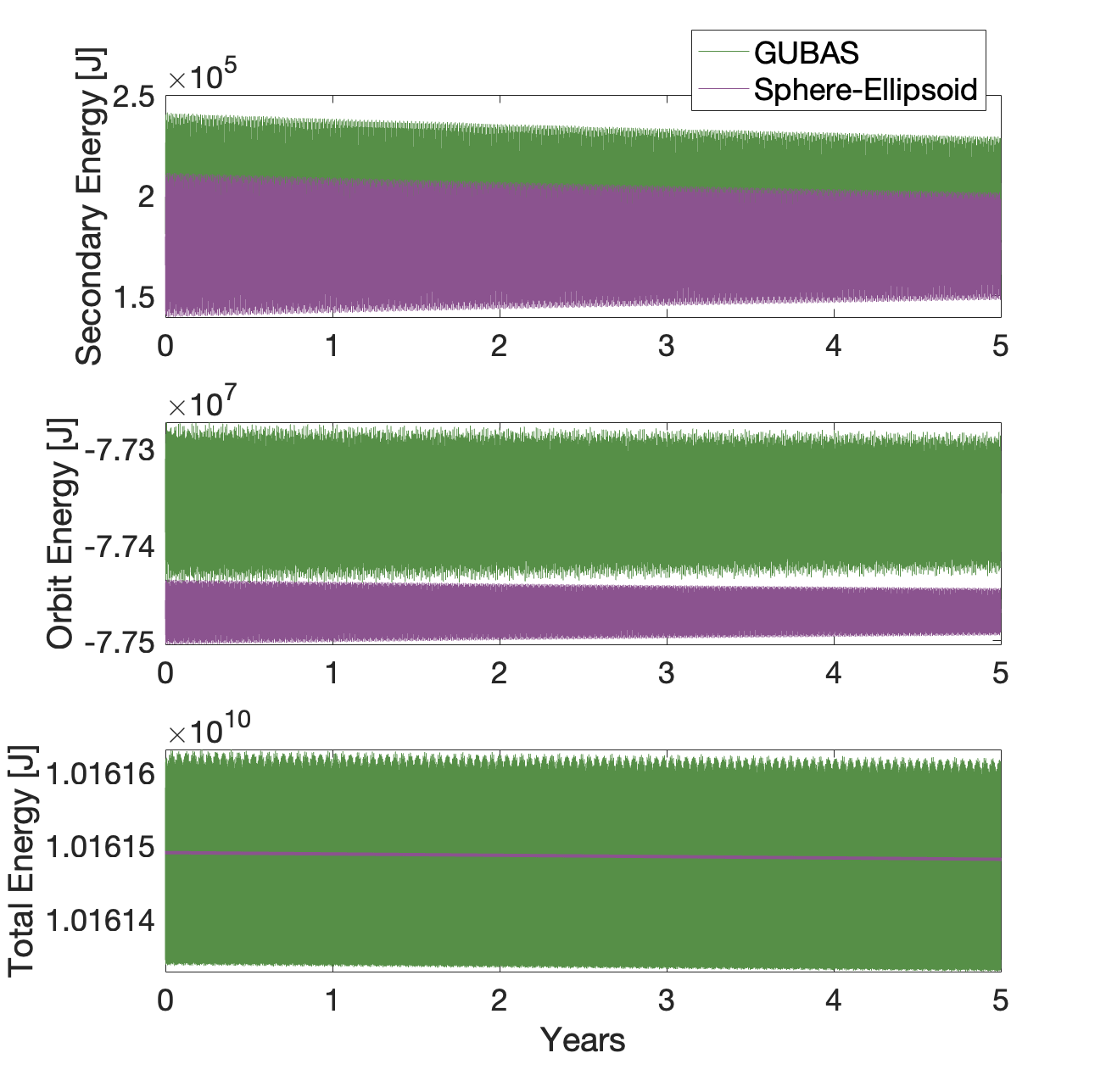} 
   \caption{The secondary energy (top), orbit energy (middle), and total energy (bottom) of the stable system ($a/b=1.2$, $b/c=1.1$) for the sphere-ellipsoid and high-fidelity \textsc{gubas} models. While there are differences between the models, the overall trends are consistent.}
   \label{model_stable_energy}
\end{figure}

\begin{figure}[ht!]
   \centering
   \includegraphics[width = 3in]{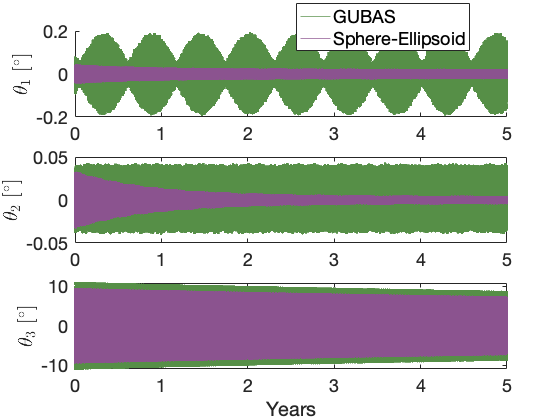} 
   \caption{The 1-2-3 Euler angles of the stable system ($a/b=1.2$, $b/c=1.1$) for the sphere-ellipsoid and high-fidelity \textsc{gubas} models. While there are differences between the models, the overall trend is consistent.}
   \label{model_stable_euler}
\end{figure}

\begin{figure}[ht!]
  \centering
  \includegraphics[width = 3in]{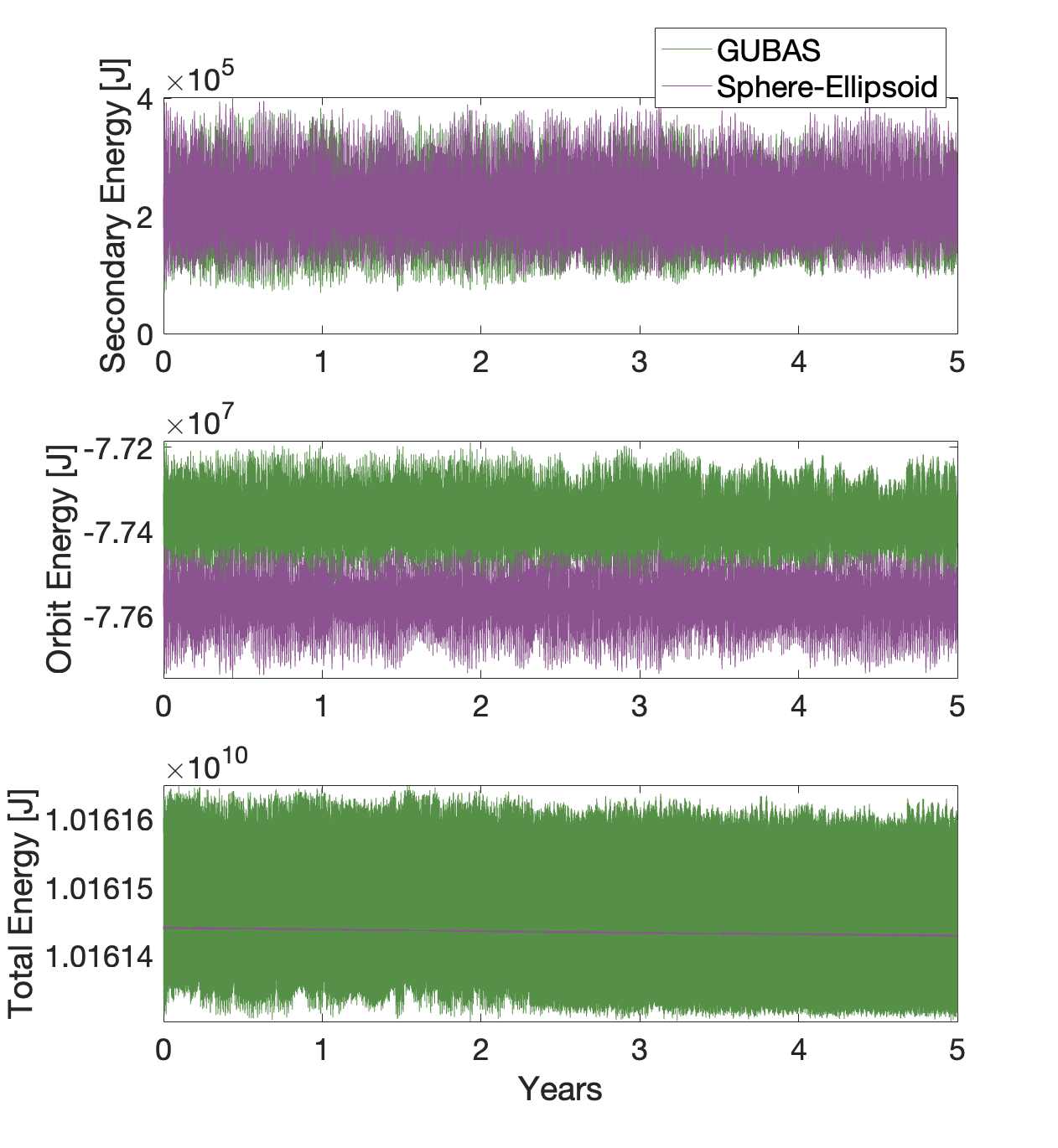} 
  \caption{The secondary energy (top), orbit energy (middle), and total energy (bottom) of the unstable system ($a/b=1.4$, $b/c=1.3$) for the sphere-ellipsoid and high-fidelity \textsc{gubas} models. While there are differences between the models, the overall trends are consistent.}
  \label{model_unstable_energy}
\end{figure}

\begin{figure}[ht!]
  \centering
  \includegraphics[width = 3in]{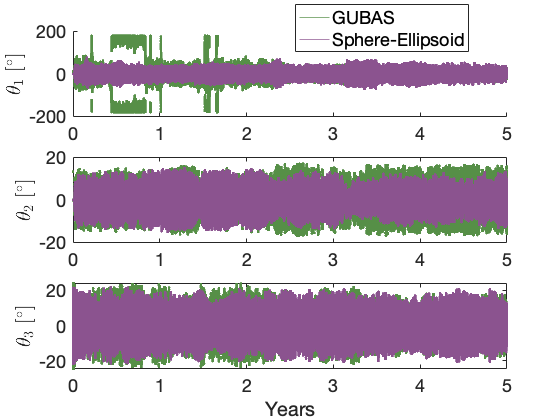} 
  \caption{The 1-2-3 Euler Angles of the unstable system ($a/b=1.4$, $b/c=1.3$) for the sphere-ellipsoid and high-fidelity \textsc{gubas} models. While there are differences between the models, the overall trends are consistent.}
  \label{model_unstable_euler}
\end{figure}

\begin{figure}[htpb!]
  \centering
  \includegraphics[width = 3in]{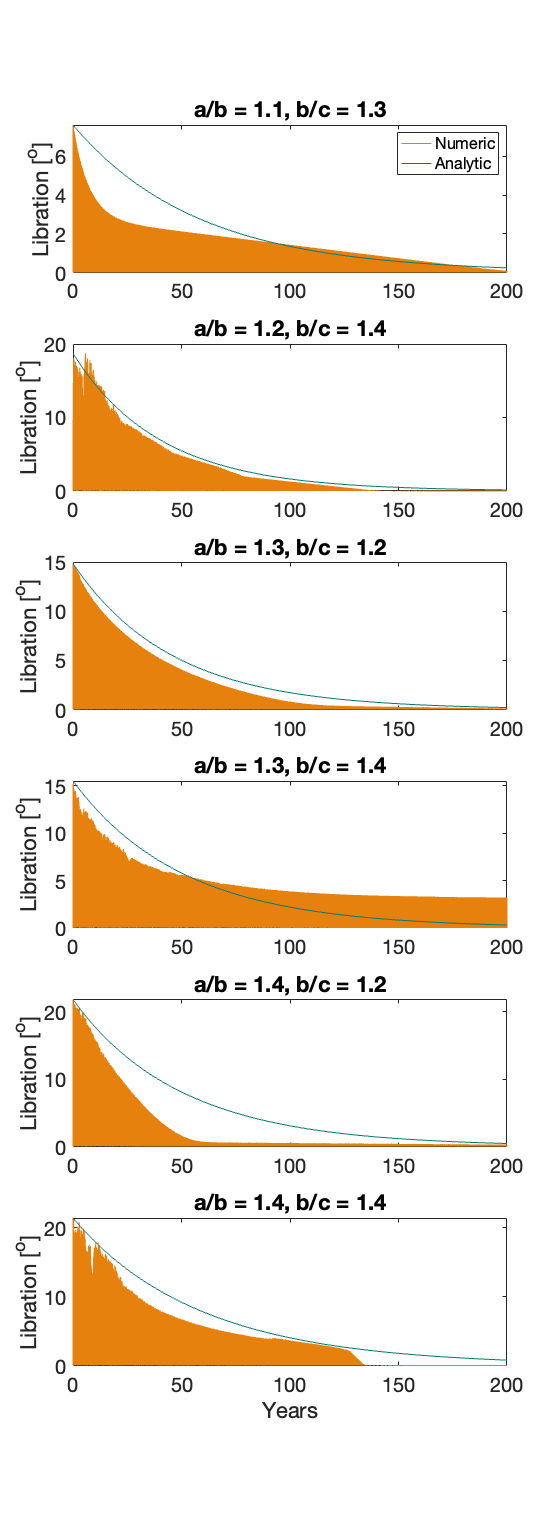} 
  \caption{The libration amplitude for 6 additional secondary shapes from both the stable and unstable regions. Every shape has a similar damping timescale regardless of stability.}
  \label{other_shapes}
\end{figure}

%\FloatBarrier
%\printcredits

%% Loading bibliography style file
%\bibliographystyle{model1-num-names}
\bibliographystyle{cas-model2-names}

% Loading bibliography database
\bibliography{bib}

%\vskip3pt

\end{document}